\documentclass[journal,comsoc]{IEEEtran}
\usepackage[letterpaper, left=1in, right=1in, bottom=1in, top=0.75in]{geometry}
\usepackage{graphicx}

\usepackage{subcaption}

\usepackage[linesnumbered, ruled, vlined, english]{algorithm2e}
\SetKwInput{kwInit}{Initialization}
\SetKwInput{kwLearning}{Learning pattern}

\usepackage{soul}
\usepackage{xcolor}
\usepackage{caption}
\usepackage{comment}
\usepackage{subcaption}
\graphicspath{ {./images/} }  
\usepackage{amsmath,amsfonts,amssymb,bbm}
\usepackage{empheq}
\newtheorem{prop}{Proposition}
\newtheorem{theorem}{Theorem}

\title{
Traffic-Aware Mean-Field Power Allocation for Ultra-Dense NB-IoT Networks
}

\author{Sami Nadif$^{*\bullet}$, Essaid Sabir$^{\bullet\circ}$, Halima Elbiaze$^\circ$ and Abdelkrim Haqiq$^{*}$\\
$^{*}$Hassan  First  University  of  Settat,  Faculty  of  Sciences  and  Techniques,  Computer, Networks, Mobility and Modeling Laboratory: IR2M, 26000 - Settat, Morocco\\
$^{\bullet}$NEST Research Group, ENSEM, Hassan II University of Casablanca, Morocco \\ $^\circ$Computer Science Department, University of Quebec at Montreal, Montreal, Canada
}

\begin{document}

\maketitle
\thispagestyle{empty}
\pagestyle{empty}

%%%%%%%%%%%%%%%%%%%%%%%%%%%%%%%%%%%%%%%%%%%%%%%%%%%%%%%%%%%%%%%%%%%%%%%%%%%%%%%%
\begin{abstract}
The Narrowband Internet of Things (NB-IoT) is a cellular technology introduced by the Third-Generation Partnership Project (3GPP) to provide connectivity to a large number of low-cost IoT devices with strict energy consumption limitations. However, in an ultra-dense small cell network employing NB-IoT technology, inter-cell interference can be a problem, raising serious concerns regarding the performance of NB-IoT, particularly in uplink transmission. Thus, a power allocation method must be established to analyze uplink performance, control and predict inter-cell interference, and avoid excessive energy waste during transmission. Unfortunately, standard power allocation techniques become inappropriate as their computational complexity grows in an ultra-dense environment. Furthermore, the performance of NB-IoT is strongly dependent on the traffic generated by IoT devices. In order to tackle these challenges, we provide a consistent and distributed uplink power allocation solution under spatiotemporal fluctuation incorporating NB-IoT features such as the number of repetitions and the data rate, as well as the IoT device's energy budget, packet size, and traffic intensity, by leveraging stochastic geometry analysis and Mean-Field Game (MFG) theory. The effectiveness of our approach is illustrated via extensive numerical analysis, and many insightful discussions are presented.
\end{abstract}


%%%%%%%%%%%%%%%%%%%%%%%%%%%%%%%%%%%%%%%%%%%%%%%%%%%%%%%%%%%%%%%%%%%%%%%%%%%%%%%%
\section{Introduction}
The Internet of Things (IoT) is one of the key applications and technologies in fifth-generation (5G) communications and beyond. Cellular IoT can be classified into four IoT connectivity segments \cite{c1}, each with distinct sets of requirements: Massive IoT, distinguished by a large number of low-cost devices, small data sizes, and restrictive energy usage requirement; Broadband IoT, characterized by large data sizes and stringent data rate requirement; Critical IoT, distinguished by ultra-reliable data delivery and stringent availability and latency requirements; And Industrial Automation IoT, characterized by Ethernet protocol integration and time-sensitive networking. Cellular networks are suitable to support all these segments, although there is no specific technology appropriate to fulfill the connectivity specifications of all potential use cases. For instance, to support massive IoT scenarios, the 3rd Generation Partnership Project (3GPP) has standardized several cellular IoT technologies, including Cat-M1 \cite{c2} and Narrow Band-IoT (NB-IoT) \cite{c3, c4, c5}. Several other works have also looked into the issue of massive connectivity in an IoT environment. In \cite{c6}, a Non Orthogonal Multiple Access (NOMA) scheme was proposed to tackle the challenge associated with connectivity and spectral efficiency in ultra-dense IoT scenarios. Choosing one or a combination of these massive IoT technologies depends on several factors, such as technology availability and coverage, use case criteria, and implementation cost. Cat-M1 technology, for example, operates at \(1.4\) MHz bandwidth to achieve higher data rates and lower latency, while LTE NB-IoT operates at \(180\) kHz bandwidth with a peak data rate of around \(250\) (resp. \(230\)) kilo-bits per second in uplink (resp. downlink) but has excellent coverage and deployment flexibility. Moreover, the NB-IoT architecture, introduced in 3GPP Rel-13 to support basic massive IoT requirements, has received several enhancements in Rel-14, such as a larger Transport Block Size (TBS) and a new device power class support of \(14\) dBm, and in Rel-15, such as small cell support \cite{c5, c7}.\\

In this article, we consider an ultra-dense small cell environment using NB-IoT technology to provide indoor connectivity for a large number of IoT devices, and we focus on ultra-low-end IoT applications with small data sizes and limited demands on throughput, such as home security, remote monitoring, and smart metering. Because of the intrinsic characteristics of these IoT applications, IoT devices must run for a long time. As a result, energy consumption becomes a critical concern. Currently, NB-IoT supports discontinuous transmission to save the IoT devices energy based on wake-up and sleep modes. However, how to further reduce their transmission energy during the wake-up phase remains an open issue. Additionally, with this massive deployment, inter-cell interference is also a key issue, especially for uplink transmission. Thus, to assess the uplink performance of the NB-IoT technology in an ultra-dense small cell network, manage and predict the inter-cell interference, and avoid unnecessary energy wastage during the wake-up period, we investigate a general spatiotemporal model that takes into account the IoT device’s energy budget, packet size, and traffic intensity, as well as the number and locations of IoT devices and Small
Base Stations (SBS). For tractability of our analysis, the SBSs are modeled using a homogeneous Poisson Point Process (PPP), and the IoT devices are clustered around them using a general cluster process. It is worth mentioning that stochastic geometry, particularly point process theory, has been widely used to model the spatial topology of cellular networks, and several empirical foundations validate the PPP model $\cite{P1,P2}$. From the temporal perspective, we consider a heterogeneous IoT traffic scenario and use Poisson process approximations based on the Palm-Khintchine theorem, which claims that the aggregated traffic from a large number of IoT devices can be approximated with a Poisson process under certain conditions. The main difficulty in considering a spatiotemporal model is that the set of active IoT devices that cause interference changes dramatically over time. Thus, to highlight the effectiveness of our approach, the power allocation problem is initially formulated as an infinite horizon dynamic differential game (due to interference coupling) whose complexity increases with the number of IoT devices. Then, we extend it to the Mean-Field Game $\cite{c8, c9, c10, c11}$, which is a powerful framework for modeling and analyzing large-scale ultra-dense networks $\cite{c12, c13, c14, c15}$. More precisely, we first introduce the mean-field regime, which describes the mass behaviors of IoT devices in a massive IoT network under partial information and symmetry assumptions. And based on that, we derive the mean-field interference, which allows each IoT device to derive its optimal power allocation strategy based only on its own energy budget, packet size, and the statistical distribution of the considered system state, known as the mean-field, while accounting for NB-IoT parameters, spatial randomness, and IoT traffic. Finally, our infinite horizon power allocation formulation leads to a stationary mean-field optimal control problem $\cite{c16,c27}$, from which we can recover a set of equations that may be solved iteratively to provide the optimal power allocation strategy.

\subsection{Related Work}
In order to fulfill the requirements of the massive IoT scenario, various studies have attempted to analyze NB-IoT power consumption and device lifetime.  The authors of \cite{f3} assessed the performance of NB-IoT in the context of a smart city. They presented a theoretical model that seeks to reduce overall energy consumption by evaluating sleep duration and the number of waking up periods. They conclude that a lifetime of 8 years is achievable under poor coverage with a reporting interval of one day. In \cite{f2}, an uplink energy-efficient and ultra-reliable heuristic algorithm has been developed to achieve high transmission reliability. More precisely, the optimization problem, which has been proven to be NP-complete, is divided into two parts. The first phase attempts to optimize the default transmit settings of IoT devices in order to incur the lowest energy consumption while meeting the quality of service requirement by considering the interference perceived at the base station as given. The second phase employs an energy-efficient weighting strategy to ensure the delay constraint per IoT device based on the scheduling emergency and inflexibility. Simultaneously, the issue of radio resource allocation and interference management cannot be neglected in the process of NB-IoT research and deployment. The authors in \cite{f4} analyzed the scenario in which just a portion of existing LTE cell sites support NB-IoT (the so-called partial deployment of NB-IoT). They investigate various strategies to compensate for the high interference, such as repetition protocol, interference mitigation techniques, and guard band deployment, and they also provide additional analysis to determine when partial NB-IoT deployment is feasible. In \cite{R4}, the authors developed a cooperative resource allocation approach for NB-IoT using cooperative interference prediction and flexible duplexing techniques. The primary idea behind cooperative interference prediction is to collaborate with a neighboring base station and exchange the expected scheduling information of IoT devices to estimate the interference level. Due to the complexity of finding the globally optimum solution, they propose a lower bound approximation by imposing a maximum interference threshold for better tractability. The work in \cite{f1} presents a resource allocation problem to optimize the uplink rate while ensuring the lowest latency computed by ignoring the scheduling and transmission time components and preserving the repetition protocol, recognized as the dominant one. They show that the optimal solution of the optimization problem is unfeasible and suggest a heuristic approach that divides the problem into two parts. The first phase handles uplink scheduling using a centralized strategy. Whereas, the second phase uses the water filling algorithm to determine the optimal power allocation for IoT devices using the same channel for transmission.

\subsection{Our Contributions}
The contributions of this paper can be summarized as follows:\\
- We develop a simple, yet insightful spatiotemporal model for ultra-dense small cells IoT networks;\\
- We present a stationary power allocation scheme that integrates spatiotemporal fluctuation, NB-IoT features, as well as the IoT device's energy budget and packet size. Our approach enables each IoT device to determine its optimal power allocation strategy based only on its own energy budget, packet size, and the mean-field; \\
- We combine stochastic geometry framework and mean-field approximation to forecast the average interference in large-scale ultra-dense IoT network while accounting for spatial randomness and IoT traffic;\\
- We propose an iterative algorithm with consistent complexity in the sense that a power allocation solution can be found after a few iterations even in an extremely dense environment (see Figure 1);\\
- We assess the performance of NB-IoT in terms of signal to interference-plus-noise ratio and transmission success probability. We also provide some fundamental insights about the repetition protocol in ultra-dense small-cell IoT networks.

The rest of the paper is organized as follows: Preliminaries on NB-IoT are presented in section II. We describe our spatiotemporal model in section III. The power allocation problem is formalized in section IV. The mean-field regime is established in section V. Extensive numerical simulations are presented in section V.  Finally, concluding remarks are given in Section VII.

\section{Preliminaries}
Ubiquitous and massive IoT systems are an attractive challenge for the emerging networks (5G and beyond). Therefore, they must handle the dramatic explosion of connected IoT devices such as machine type communication devices (automated devices that require minimum human intervention). To support such a massive connection and improve the battery life of low data rate IoT devices, a Narrow Band IoT (NB-IoT) system has been developed by the 3rd generation partnership project as the communication standard for IoT. The NB-IoT uses Single Carrier Frequency Division Multiple Access (SC-FDMA) for uplink communication with a Narrowband Physical Uplink Shared Channel in Format 1 (NPUSCH-F1) for carrying uplink data.
\subsection{NB-IoT Deployment Modes }
To improve spectrum usage and decrease the deployment cost, NB-IoT offers three deployment modes by reusing the existing spectrum of LTE and GSM:
\begin{itemize}
\item Stand-alone mode: uses separate frequency bands (GSM carrier) of 200 kHz bandwidth and does not interfere with the LTE frequency band.
\item In-band mode: occupies a total system bandwidth of 180 kHz that corresponds to one LTE Physical Resource Block (PRB) inside the LTE carrier.
\item Guard-band mode: uses one PRB bandwidth as an access spectrum in the LTE edge frequency band.
\end{itemize}

\subsection{Resource Unit }
To allow a bandwidth allocation smaller than one LTE PRB in uplink communication, NB-IoT relies on Resource Units (RUs). They represent the minimal dispatching units allocated to IoT devices for NPUSCH-F1 transmission. Thus, the transmission data (packet) of an IoT device is carried by one or multiple RUs, \(N^{ru}\), depending on the packet size and the Modulation and Coding Scheme (MCS) level. Besides, the NB-IoT resource grid employs 3.75 or 15 kHz subcarrier spacing, resulting in multiple RU categories, each with a different number of subcarriers and time slots. These subcarriers are called tones. Multi-tone transmission uses the 15 kHz subcarrier spacing grouped into 3, 6, or 12 continuous tones with a 0.5 ms slot duration. However, single-tone transmission adopts either 3.75 or 15 kHz numerology with a 2 ms or 0.5 ms slot duration, respectively.

\begin{table}
\centering 
\caption{Resource Units (RUs)}
\begin{tabular}{|p{1.3 cm}|p{1.8 cm}|p{1.5cm}|p{1.2 cm}|}
  \hline
 \centering  \textbf{Tone bandwidth} & \centering \textbf{Number of tones per RU}&  \textbf{Number of slots per RU} & \centering \textbf{slot} \par\textbf{ duration }\tabularnewline 
  \hline
   \centering{3.75 kHz} & \centering{1} & \centering{16} & \centering 2 ms \tabularnewline 
  \hline
              & \centering 1 & \centering{16} & \tabularnewline 
  \cline{2-3}
 \centering{15 kHz}  &\centering 3 & \centering8 & \centering 0.5 ms \tabularnewline 
  \cline{2-3}
              & \centering6 & \centering4 & \tabularnewline 
   \cline{2-3}
              &\centering {12} & \centering2 & \tabularnewline 
  \hline
\end{tabular}
\end{table}

\subsection{Repetition Protocol}
To boost demodulation efficiency and enlarge the network coverage, the NB-IoT adopts a repetition protocol. According to the NB-IoT norm, the assigned resource units of each IoT device are associated with a particular number of repetitions \(N^{rep}\). Indeed, to fulfill the coverage requirements of different IoT devices, NB-IoT networks can configure up to \(3\) repetition values from the set \(\{1, 2, 4, 8, 16, 32, 64, 128\} \) in a cell, and allows flexible configuration of NPSUCH-F1 resources. However, even if NB-IoT targets latency-insensitive applications, it is designed to allow less than 10 s latency and also aims to support a long battery life. Thus, there is a trade-off between reliability, latency, and energy consumption.

\begin{table}
\centering 
\caption{Transport block size (TBS)}
\small\addtolength{\tabcolsep}{-2.5pt}
\begin{tabular}{|c|c|c|c|c|c|c|c|c|}
  \hline
  \textbf{MCS}& \multicolumn{8}{c|}{\textbf{Number of resource units (\(N^{ru}\))}}\tabularnewline 
   \cline{2-9}
 \textbf{Level}& \textbf{1}&\textbf{ 2}&\textbf{ 3}&\textbf{4}& \textbf{5}&\textbf{ 6}& \textbf{8}& \textbf{10}\tabularnewline 
  \hline
   \textbf{0}& 16& 32 & 58 &88&120&152&208&256\tabularnewline 
  \hline
     \textbf{1}& 24& 56 & 88 &144&176&208&256&344\tabularnewline 
\hline
     \textbf{2}& 32& 72 & 144 &176&208&256&328&424\tabularnewline 
  \hline
\textbf{3} &40 &104& 176& 208& 256& 328& 440& 568 \tabularnewline 
\hline
\textbf{4 }&56& 120& 208& 256& 328& 408& 552& 680 \tabularnewline 
\hline
\textbf{5 }&72 &144 &224 &328 &424 &504 &680 &872 \tabularnewline 
\hline
\textbf{6} &88 &176 &256 &392 &504 &600& 808& 1000 \tabularnewline 
\hline
\textbf{7 }&104 &224 &328 &472 &584 &712 &1000 &1224 \tabularnewline 
\hline
\textbf{8} &120 &256 &392 &536 &680 &808 &1096 &1384 \tabularnewline 
\hline
\textbf{9} &136 &296 &456 &616 &776 &936 &1256 &1544 \tabularnewline 
\hline
\textbf{10} &144 &328 &504 &680 &872 &1000 &1384 &1736 \tabularnewline 
\hline
\textbf{11} &176 &376 &584 &776 &1000 &1192 &1608 &2024 \tabularnewline 
\hline
\textbf{12} &208 &440 &680 &1000& 1128 &1352& 1800 &2280 \tabularnewline 
\hline
\textbf{13 }&224 &488 &744 &1032 &1256 &1544 &2024 & 2536 \tabularnewline 
\hline
\end{tabular}
\end{table}

\subsection{Data Rate}
In this subsection, we derive the uplink data rate based on the information presented in the previous subsections. In NB-IoT, the data rate is increased by reducing the transmission time, i.e., by allocating fewer RUs for a given TBS. This improves the IoT device’s energy usage but lowers the coverage area at the same time. Coverage extension is achieved with the help of repetitions instead. However, in some circumstances, the data rate may be intentionally reduced to improve the coding gain, and in most cases, the energy per information bit as well. Therefore, in order to model the uplink data rate of NB-IoT, the above mentioned factors need to be considered. \\
NB-IoT supports MCS levels from 0 to 13, using QPSK or BPSK modulation, and a variable TBS up to 2536 bits (3GPP Rel-14). The TBSs supported for NPUSCH Format 1 are shown in Table II. Note that the TBSs in the final three rows of Table II are exclusively used for multi-tone transmissions. Moreover, each packet is divided into several Transport Blocks (TBs), and each TB is mapped into several RUs. As mentioned before, the RU duration, denoted as \(T^{ru}\), depends on the number of tones and time slots used. Thus, the transmission time of a specific packet is given as follows:
\begin{equation}
T^{tr} =   \bigg\lceil \frac{PacketSize}{TBS} \bigg\rceil \times N^{ru}  \times T^{ru}\times N^{rep},
\label{time_tr}
\end{equation}
where \(\lceil .\rceil\) represents the ceiling operator. It can be easily seen that the transmission time per IoT device depends on the total number of transport blocks needed to transmit the IoT packet, i.e., \(\bigg\lceil \frac{PacketSize}{TBS} \bigg\rceil\), the time needed to transmit one transport block, i.e., \( N^{ru}  \times T^{ru}\), and the number of repetitions.\\
Accordingly, the data (transmission) rate of a specific packet can be calculated as:
\begin{equation}
R^{tr} = \frac{PacketSize}{T^{tr}} = \frac{PacketSize}{ \bigg\lceil \frac{PacketSize}{TBS} \bigg\rceil \times N^{ru}  \times T^{ru}\times N^{rep}}
\label{rate}
\end{equation}
\begin{table}[h]
\centering 
\caption{NB-IoT uplink peak data rate in different
deployment scenarios.}
\begin{tabular}{|p{2.8 cm}|p{0.5 cm}|p{0.5cm}|p{0.5 cm}|p{0.5cm}|p{0.5cm}|}
  \hline
 \centering  Configuration& \multicolumn{2}{c|}{Single-tone} & \multicolumn{3}{c|}{Multi-tone}\tabularnewline 
  \hline
 \centering  Tone bandwidth (kHz) & \centering 3.75& \centering 15& \multicolumn{3}{c|}{15}\tabularnewline 
  \hline
 \centering  Number of tones& \centering 1&  \centering 1& \centering 3 & \centering 6 & \centering 12\tabularnewline 
 \hline
  \centering  \textbf{Peak data rate (kbps)}& \centering \textbf{8}&  \centering\textbf{32}& \centering \textbf{64} & \centering\textbf{129} & \centering\textbf{258}\tabularnewline 
  \hline
\end{tabular}
\end{table}
\section{System Model}
In this paper, we focus on a NB-IoT system with a total bandwidth of \(180\) kHz using the single-tone configuration (either the \(3.75\) or \(15\) kHz tone bandwidth). Additionally, we assume synchronous NB-IoT deployment in all small cells using the same LTE PRB. It means that the PRB utilized for NB-IoT communication is the same across the entire network, and the source of interference is only caused by the neighboring small cells (inter-cell interference).

\subsection{Spatial Model}
We consider an ultra-dense cellular network in which the SBSs are spatially distributed in the euclidean plane according to a homogeneous PPP, denoted as \(\phi_s=(\textbf{z}_1,\textbf{z}_2,...)\), with intensity \(\beta_s\). Each SBS uses \(c = TotalBandwith/ToneBandwith\) different tones to provide connectivity for a large number of IoT devices, which are assumed to operate mainly in the uplink and belong to a given closed subscriber group. \\
The IoT devices are modeled using a cluster process, where the cluster centers are the SBSs. The number of IoT devices associated with a specific SBS at the location \(\textbf{z} \in \phi_s\), defined as \(N_{\textbf{z}}\), follows a Poisson distribution with  parameter \(\beta_u\):
\begin{equation}
\mathbb{P}[N_{\textbf{z}}=k]=e^{-\beta_u }\dfrac{(\beta_u )^k}{k!}, \quad k=0,1,\dots  \quad .
\end{equation}
Let $\mathcal{C}_\textbf{z}$ be the coverage area (disk of center \textbf{z} and radius $R_s$) of the SBS at the location $\textbf{z}$. We denote by \(\phi_{\textbf{z}}=(\textbf{y}_1, \textbf{y}_2, \dots )\) the point process of IoT devices associated to SBS at the location \(\textbf{z}\) whose locations are independently and identically distributed with a probability density function $
f(\cdot|\textbf{z})$. Moreover, by using the polar coordinate system, we write
\begin{equation}
f(\textbf{y}|\textbf{z})=\frac{f_r( \mid\mid \textbf{y}-\textbf{z}\mid\mid)}{2\pi \mid\mid \textbf{y}-\textbf{z} \mid\mid}, \quad \textbf{y} \in \mathcal{C}_{\textbf{z}}/\{\textbf{z}\},
\end{equation}
where $\mid\mid \cdot \mid\mid $ denotes the euclidean distance and $f_r(\cdot)$ is a probability density function on $]0,R_s]$.\\
In this work, the distance of an arbitrary IoT device from its serving SBS follows a general probability density function, such that \(\frac{r}{R_s}\) follows a Beta\((a,b)\) distribution, expressed as:
\begin{equation}
f_r(r)=\frac{1}{R_s}\frac{(\frac{r}{R_s})^{a-1}(1-\frac{r}{R_s})^{b-1}}{\int_0^1u^{a-1}(1-u)^{b-1}du}, \quad 0 < r \leq R_s.
\label{beta_dist}
\end{equation}
This density allows us to consider different cases of IoT device distribution per small cell, based on the parameters \(a\) and \(b\):\\
\(\bullet\) If \(a>1\) and \(b>1\), the IoT devices are away from both the SBS and the small cell edges.\\
\(\bullet\) If \(a \geq 1\), \(0<b\leq1\) and \((a,b)\neq(1,1)\), the IoT devices are away from the SBS and are around the small cell edges. In addition, if \((a,b) = (2,1)\), then, in this case, our cluster process is a Matern cluster process. \\
\(\bullet\) If \(0<a\leq1\), \(b \geq 1\) and \((a,b)\neq(1,1)\), the IoT devices are around the SBS and are away from the small cell edges. \\
\(\bullet\) If \(0<a<1\) and \(0<b<1\), the IoT devices are around both the SBS and the small cell edges.\\
\(\bullet\) If \(a=1\) and \(b=1\), the IoT devices are distributed uniformly in the circle of radius \(R_s\).\\
Note that, the number and locations of SBSs and IoT devices are fixed all time once they are deployed, and the resulting cluster process $\phi_u = \bigcup_{\textbf{z} \in \phi_s} \phi_{\textbf{z}}$ is a stationary point process with intensity $\beta_s \beta_u $.\\
In the rest of this paper, we consider a large circle of radius \(R\gg R_s\) to be the spatial domain of our analysis, denoted as \(\mathcal{C}_R\), and we will focus, without loss of generality, on the active IoT device using a specific tone for transmission, denoted by \(J\). Let us denote by \(P_a(t,\textbf{z})\), the probability that a SBS at the location \(\textbf{z}\) has an active IoT device using the tone \(J\) for transmission at time \(t\). We apply a spatially dependent \(P_{a}(t,\textbf{z})\)-thinning to the PPP $\phi_{s}$, then the retained SBSs form a non homogeneous PPP, denoted as $\phi_{s,t}^{a}$, with intensity \(\beta_s P_a(t,\textbf{z})\) \cite{t1}. Thus, the point process consisting of active IoT devices using the same tone $J$ for transmission at time $t$, denoted as $\phi_{u,t}^{a}$, has an intensity function conditioned on $\phi_{s,t}^{a}$ given by:
\begin{equation}
\Lambda_{\phi_{u,t}^{a} | \phi_{s,t}^{a}}(\textbf{y}) = \sum_{\textbf{z}\in \phi_{s,t}^{a}}  f(\textbf{y}|\textbf{z}) \mathbf{1}_{\{\textbf{y} \in \mathcal{C}_{\textbf{z}}\}},
\end{equation}
Therefore, by using Campbell's formula $\cite{t1}$, the (unconditional) intensity function of $\phi_{u,t}^a$ is given by:

\begin{equation}
\begin{split}
\Lambda_t(\textbf{y})&=\mathbb{E}_{\phi_{s,t}^{a}}\left[\sum_{\textbf{z}\in \phi_{s,t}^{a}}  f(\textbf{y}|\textbf{z}) \mathbf{1}_{\{\textbf{y} \in \mathcal{C}_{\textbf{z}}\}}\right]\\
&=\beta_s \int_{\mathbb{R}^2}   P_{a}(t,\textbf{z}) f(\textbf{y}|\textbf{z}) \mathbf{1}_{\{\textbf{y} \in \mathcal{C}_{\textbf{z}}\}} d\textbf{z}\\
&=\beta_s \int_{\mathbb{R}^2}   P_{a}(t,\textbf{z}) f(\textbf{y}|\textbf{z}) \mathbf{1}_{\{\textbf{z} \in \mathcal{C}_{\textbf{y}}\}} d\textbf{z}\\
&=\beta_s \int_{ \mathcal{C}_{\textbf{y} } }  P_{a}(t,\textbf{z}) f(\textbf{y}|\textbf{z}) d\textbf{z}.
\end{split}
\end{equation}
Moreover, in the special case where $P_a$ is spatially independent, the intensity function $\Lambda_{t}(\cdot)$ can be expressed as follows:

\begin{equation}
\begin{split}
\Lambda_t(\textbf{y}) &=\beta_s P_a(t) \int_{ \mathcal{C}_{\textbf{y} } }  \frac{f_r( \mid\mid \textbf{y}-\textbf{z}\mid\mid)}{2\pi \mid\mid \textbf{y}-\textbf{z} \mid\mid} d\textbf{z} \\
 &=\beta_s P_a(t) \int_{0}^{R_s}  f_r(r) dr \\
  &=\beta_s P_a(t) \\
\end{split}
\end{equation}
Finally, It is worth mentioning that the number of active IoT devices using the same tone \(J\) for transmission at time \(t\) in \(\mathcal{C}_R\), denoted as \(\mathcal{N}_{R,t}\),  is a Poisson random variable with mean intensity $\int_{\mathcal{C}_R}\Lambda_t(\textbf{y})d\textbf{y}$.

\subsection{Transmission Model}
Each IoT device \(k\) receives at random times a packet \(j\) of random size \(B_{k,j,0}\) that must be communicated to its associated SBS, as well as a certain energy budget \(E_{k,j,0}\) for transmitting those bits. Since each IoT device sends one packet at a time, thus with a slight abuse of notation, we note \(B_{k,0} := B_{k,j,0}\) and \(E_{k,0} := E_{k,j,0}\). Henceforth, we will dismiss the packet's index in the rest of this paper. To evaluate this transmission, we rely on the received time varying Signal to Interference-plus-Noise Ratio (SINR) for a Gaussian single input single output channel with the interference treated as noise:
\begin{equation}
\gamma_k(t,p_k,\textbf{p}_{-k})=\dfrac{p_k(t)H^2_{k,k}(t)L_{k,k}(r)}{\mathrm{N}_0B_w+I_k(t,\textbf{p}_{-k})},
\end{equation}
where in the above expression, \(p_k \in [0, P_{max}]\) is the transmit power of IoT device \(k\), \(\textbf{p}_{-\textbf{k}}\) denotes the transmit power vector of the active IoT devices using the same tone without \(k\), \(H_{k,j}\) (resp. \(L_{k,j}\)) is a parameter representing the multipath fading (resp. path loss) between the IoT device \(j\) and SBS serving the IoT device \(k\), \(\mathrm{N}_0\) is the noise power spectral density and \(I_k(t, \textbf{p}_{-k})\) is the inter-cell interference expressed as:
\begin{equation}
I_k(t, \textbf{p}_{-k}) = \sum_{j=1 ,  j \neq k}^{|\mathcal{N}_{R,t}|} p_j(t)H^2_{k,j}(t)L_{k,j}(r).
\end{equation}
In our analysis, the path-loss function \(L : \mathbb{R}^+ \to \mathbb{R}^+\), which encodes how the signal power attenuates with distance, follows a power-law model defined as \(L(r)  = L_0 r^{-\alpha(r)}\), where \(\alpha\) is the path-loss exponent (changes with distance), and \(L_0\) is the path-loss at \(r=1\) m. On a logarithmic scale, this corresponds to a straight line path loss with a slope of \(10 \alpha(r) \log_{10}(r)\). Furthermore, a distance-dependent exponent that increases from \(2\) to \(12\) has been reported for indoor measurements carried out in a multistory office building \cite{c17}. The large values of \(\alpha\) are due to an increase in the number of walls and partitions between the transmitter and receiver when the distance increases. In this work, we follow similar approach such as the values of \(\alpha\) are \(\alpha_1\) (for \(0 < r \leq 3\) m), \(\alpha_2\) (for \(3 < r  \leq 20\) m), \(\alpha_3 \) (for \(20< r \leq 40\) m), and \(\alpha_4 \) (for \(r > 40 \) m) with \(\alpha_1 \leq \alpha_2 \leq \alpha_3 \leq\alpha_4\). Another approach consists of associating logarithmic attenuation with the number of intervening walls between the transmitter and receiver. Adding these individual attenuations, the total path loss can be calculated \cite{c18}. \\
Additionally, the multipath fading \(H\), which represents the signal variations due to multipath propagation, follows a Rician distribution \(Rice(\eta,\sigma)\), where \(\eta^2\) is the power of the direct line of sight or the dominant multipath component, and \(\sigma^2\) is the power of the rest multipath component \cite{c19}. We assume that the transmitted signals between an IoT device and its serving (resp. non-serving) SBS experience independent and identically distributed Rician fading \(Rice(\eta,\sigma_1)\) (resp. Rayleigh fading \( Rayleigh(\sigma_2) = Rice(0,\sigma_2)\) ) in order to model an indoor environment (resp. outdoor environment).
\subsection{IoT Traffic}
We model the arrival of new packets at the IoT device \(k\) as an independent renewal process with the arrival intensity \(\lambda_k\). Each device is considered to have an unlimited buffer, where arriving packets are kept in the buffer until successful transmission. Furthermore, each IoT device transmits packets using a First In First Out (FIFO) policy, which treats all packets equally by placing them at the end of the queue as they arrive.\\
The essential theorem for the superposition of traffic processes is the Palm-Khintchine theorem which claims that the superposition of a large number of independent renewal processes, each with a limited intensity, behaves asymptotically like a Poisson process.\\
\begin{theorem}[Palm–Khintchine theorem]
Let us consider \(n\) independent renewal processes with independent and identically distributed inter-arrival times \(S_k\), \(k=1,\dots,n\). The expected inter-arrival time for each process is \( \mathbb{E}[S_k]=1/\lambda_k\) where \(\lambda_k\) is the arrival intensity. If the following assumptions hold:\\
1) \(\lambda_1 + \dots + \lambda_n = \lambda < \infty\) when \(n \to \infty\).\\
2) Given \(\epsilon > 0\), for every \(t>0\) and when \(n \to \infty\): \(\mathbb{P}(S_k \leq t) < \epsilon\) for all \(k\).\\
Then the superposition of the renewal processes approaches a Poisson process with intensity \(\lambda\) as \(n \to \infty\).
\end{theorem}
\(\quad\)\\
The implication of the Palm–Khintchine theorem for our IoT system model is as follows:\\
- SBS level (traffic per small cell) : The superposition of the renewal processes of IoT devices associated to the SBS at the location \(\textbf{z}\) can potentially be approximated by a Poisson process with intensity \(\lambda_s(\textbf{z}) = \sum_{k=1}^{N_\textbf{z}} \lambda_k\) when \(N_\textbf{z}\) is large. \\
- Large scale level (traffic in the network) : The superposition of the renewal processes of IoT devices located in \(C_R\) can potentially be approximated by a Poisson process with intensity \(\lambda = \sum_{\textbf{z} \in \phi_s(R)}\lambda_s(\textbf{z})\). \\
This is because it is fair to presume that the IoT devices emit packets independently of one another and at the same sampling frequency. Note that the Palm-Khintchine theorem makes no further assumptions regarding the individual renewal process, which might comprise both periodic and event-based processes. We refer the reader to \cite{R6} for a better understanding of the accuracy of this Poisson process approximation based on the Palm-Khintchine theorem.
\section{Problem Formulation}
In this section, we model the uplink power allocation problem for the spatiotemporal system model described above using an infinite horizon dynamic differential game. We employ a framework of non-cooperative games to devise a fully distributed algorithm. The goal of each IoT device \(k\) when it receives a packet at a random time is to determine the optimal power allocation strategy to use for transmission while minimizing an average utility function under spatiotemporal fluctuation. The state of an active IoT device \(k\) (using tone $J$ for transmission), also to be called player, at time \(t\) is described by $\textbf{X}_{k}(t) :=\textbf{X}_{k,t}=[E_{k,t}, B_{k,t}]^{T}$ and has the following dynamic:
\begin{equation}
d\textbf{X}_{k,t}=
\begin{bmatrix}
-p_k(t)\\
-R_k^{tr}
\end{bmatrix}
dt, 
\label{state}
\end{equation}
where $E_{k,t}$ is the remaining energy budget at time $t$, $B_{k,t}$ is the remaining bits in the packet at time $t$, and \(R_k^{tr}\) is the data rate for the \(k^{th}\) IoT device defined as in (\ref{rate}) by:
\begin{equation}
R^{tr}_k  = \frac{B_{k,0}}{ \bigg\lceil \frac{B_{k,0}}{TBS_k} \bigg\rceil \times N^{ru}_k  \times T^{ru}_k\times N^{rep}_k}.
\end{equation}
In this dynamic differential game, the control of an active IoT device $k$ is its transmit power $p_k(t)$ which is allowed to depend not only on time, but on its own state $\textbf{X}_{k,t}$ and on the states of all other active IoT devices in the system at time $t$, denoted as $\textbf{X}_{-k,t}$. Thus, the power allocation strategy of the player $k$ will be denoted by $p_k$ with $p_k(t) = p_k(t,\textbf{X}_{k,t}, \textbf{X}_{-k,t})$. \\
Note that in the state dynamic ($\ref{state}$), the term $dE_{k,t} = -p_k(t)dt$  implies that the variation in the energy budget during $dt$ is proportional to the transmission power. Each IoT device can select an energy budget based on several factors such as battery life, location, IoT traffic, and so on, to fully utilize it during transmission. This is especially important for IoT devices that must operate without human intervention over an extended period of time. The term $dB_{k,t} =-R_k^{tr}dt$, on the other hand, denotes that the number of bits transmitted during $dt$ is proportional to the data rate. Furthermore, any IoT device \(k\) that becomes active at time \(T^a_k\) with a packet of size \(B_{k,0}\) and an energy budget \( E_{k,0}\) will stop transmitting ($p_{k}(t) = 0$) and exit the game once it has sent all the information (\(B_{k,t} = 0\)) or when it runs out of energy (\(E_{k,t} = 0\)). We denote by \(\theta_k\) (the exit time) the first time when one of these conditions is satisfied. It is worth mentioning that the exit time is not a control, but rather a function of the IoT device's whole trajectory, and when the condition \(B_{k,t} = 0\) comes first, it is equal to the transmission time defined in (\ref{time_tr}).\\
The goal of each IoT device is to transmit its packet within its energy budget while maximizing its throughput. Thus, the average utility function of a player \(k\) is given by:
\begin{equation}
\begin{split}
U_k(p_k,\textbf{p}_{-k}) & = \mathbb{E}\left [\int\limits_{0}^{\infty}-F_k(t, p_k, \textbf{p}_{-k}) dt  \right]\\
 & = \mathbb{E}\left [\int\limits_{T^a_k}^{T^a_k+\theta_k}-F_k(t, p_k, \textbf{p}_{-k}) dt  \right],
\end{split}
\end{equation}
where \(F_k := \log_2(1 + \gamma_k)\) is proportional to the Shannon channel capacity.\\ 
Taking the packet dynamic into account in ($\ref{state}$) will change the outcome of the game by allowing the IoT devices to make the best use of their energy budget and ensuring that the average utility function $U_k$ is finite ($\theta_k  \leq$  $T_k^{tr}$). Therefore, a power allocation strategy profile $\textbf{p}^* = (p_1^*, p_2^*, \dots)$ is a feedback Nash equilibrium of the dynamic differential game if and only if $\forall k, p_k^*$ is a solution of the following optimal control problem:
\begin{equation}
\inf_{ p_k } U_k(p_k,\textbf{p}^*_{-k}),
\label{Opti_p}
\end{equation}
subject to
\begin{equation}
d[\textbf{X}_{k,t},\textbf{X}_{-k,t}]^{T}=
\begin{bmatrix}
-[p_k(t),\textbf{p}_{-k}^*(t)]^T\\
-\textbf{R}^{tr}
\end{bmatrix}
dt, 
\label{system_state}
\end{equation}
where $\textbf{R}^{tr} = [R^{tr}_1, R^{tr}_2, \dots ]^T$.\\
The IoT devices can enter and exit the game several times as long as they get some packets to transmit. As a result, the number of players in the game does not decrease with time, preventing the game from becoming trivial. Note that when an IoT device re-enters the game, it is treated as a new player.\\
For the above formulation, as the IoT devices' behavior changes with time, the transmit power of each player must adapt accordingly. Therefore, to obtain the optimal power allocation strategies, the standard solution concept consists of analyzing the Nash equilibrium. However, the complexity of this approach increases with the number of players. Additionally, it requires that each player have complete knowledge of the states and actions of all the other players, which brings about a large amount of information exchange. This is not feasible and impractical for an ultra-dense NB-IoT system. Nevertheless, since the effect of other players on a single player’s average utility function is only via inter-cell interference, it is intuitive that, as the number of players increases, a single player has a negligible effect. Thus, we suggest using a mean-field limit for this game to convert these multiple interactions (inter-cell interference) into a single aggregate interaction known as mean-field interference.

\section{Mean Field Game}
In this section, we first introduce the mean-field concept, and based on it, we derive the mean-field interference and then formulate the stationary mean-field optimal control problem.
\subsection{Mean Field Regime}
The general setting of a mean-field regime is based on the following assumptions:
\begin{itemize}
\item The existence of large number of players ensured by considering large scale ultra dense network.
\item Interchangeability: the permutation of the state among the players would not affect the optimal power allocation strategy. 
\item Finite mean-field interference \(I_{mf}\).
\end{itemize}
To ensure the interchangeability property, the players should only know their individual states, have a homogeneous transmit power \(p_{k}(t)=p(t,\textbf{X}_{k,t})=p(t,E_{k,t},B_{k,t})\) and must have the same transmission rate, i.e, \(R_{1}^{tr} = R_{2}^{tr} = \dots = R^{tr}\). Thus, we consider single repetition value to provide fundamental insights about the optimal power allocation strategy, assume that the players use the same MCS level and the same number of resource units, and consider the following transmission rate approximation (upper bound):
\begin{equation}
R^{tr}  = \frac{TBS}{ N^{ru}  \times T^{ru}\times N^{rep}}.
\end{equation}
All the results obtained here can be generalized to the case of unequal rates and different repetition values by considering a mean-field game with different populations \cite{c20}.\\

We first consider a time interval \([0,T]\)  where we suppose that there is no arrival of new players. The players in the system are fixed and do not increase in \([0,T]\). Let's note \(\Omega= [0,E_{max}]\times [0,B_{max}]\) to be the state domain of our analysis. We define the empirical state distribution of the players in \(\mathcal{C}_R\) at time \(t\) in \([0,T]\) as:
\begin{equation}
\ M(t,\textbf{x})= \frac{1}{|\mathcal{N}_{R,0}|}\sum_{k=1}^{|\mathcal{N}_{R,0}|}\delta_\textbf{x}(\textbf{X}_{k,t}), \quad \forall \textbf{x} = (e,b) \in \Omega,
\label{MF}
\end{equation} 
where \(\delta_\textbf{x}\) is the Dirac measure.\\
The basic idea behind the mean-field regime is to approximate a finite population with an infinite one, where the empirical state distribution \(M(t,\textbf{x})\) almost surely converges, as \(|\mathcal{N}_{R,0}| \to \infty\), to the probability density function \(m(t,\textbf{x})\) of a single player, due to the strong law of large numbers. We will refer to the probability density function \((m_t)_{t\geq 0}\) as the mean-field. Additionally, as the players become essentially indistinguishable, we can focus on a generic player by dropping his index \(k\) where his individual dynamic is written as: 
\begin{equation}
d\textbf{X}_{t}=
\begin{bmatrix}
-p(t, \textbf{X}_t)\\
-R^{tr}
\end{bmatrix}
dt.
\end{equation}
To derive the Fokker Planck Kolmogorov (FPK) equation which modelled the evolution of the mean-field \((m_t)_{t\geq 0}\) over time, we need to be able to express the average values of any deterministic smooth function \(\phi(\textbf{X}_t)\) in terms of \(m(t,\textbf{x})\). It is obvious that the result can be represented as an average with respect to the distribution of \(\textbf{X}_t\), i.e.,
\begin{equation}
\mathbb{E}[\phi(\textbf{X}_t)] = \int_{\Omega}\phi(\textbf{x})m(t,\textbf{x})d\textbf{x}.
\end{equation}
By using Dynkin's formula, we have for any deterministic smooth function \(\phi : \Omega \to \mathbb{R} \):
\begin{equation}
\begin{split}
& \mathbb{E}\left [\phi(\textbf{X}_t) - \phi(\textbf{X}_0) \right ] = \\
&  \mathbb{E} \left [\int_0^t \left (-p(s,\textbf{X}_s)\partial_e \phi(\textbf{X}_s) - R^{tr}\partial_b\phi(\textbf{X}_s) \right ) ds \right ]. 
\end{split}
\label{dynkin}
\end{equation}
 Then (\ref{dynkin}) can be rewritten as
 \begin{equation}
 \begin{split}
 & \int_{\Omega}( m(t,\textbf{x}) - m(0,\textbf{x}))\phi(\textbf{x})d\textbf{x} = \\
 & \int_0^t \int_{\Omega}\big (-p(s,\textbf{x}_s)\partial_e \phi(\textbf{x}_s) - R^{tr}\partial_b\phi(\textbf{x}_s) \big) m(s,\textbf{x})  d\textbf{x}ds.
 \end{split}
 \end{equation}
Finally, the Integration by parts yields to FPK equation:
\begin{equation}
\partial_tm(t,\textbf{x})-\partial_{e}(pm)(t,\textbf{x}) -  R^{tr}\partial_bm(t,\textbf{x})=0.
\end{equation}
Now, we turn our attention to determining a utility function which depends only on the individual transmit power of a generic player and the mean-field.\\
For a generic player, we define the mean-field SINR as:
\begin{equation}
\gamma(t, p, I_{mf}) = \cfrac{p(t,\textbf{X}_t) \mathbb{E}[H^2_{indoor}] L(\mathbb{E}[r]) }{N_0B_w+I_{mf}(t)}.
\label{MF_sinr}
\end{equation}
For simplicity of our analysis, we choose \(\sigma_1 = 1\). Since \(H_{indoor} \sim Rice(\eta,\sigma_1)\), thus, \(H^2_{indoor}\) has a noncentral chi-squared distribution with two degrees of freedom and noncentrality parameter \(\eta^2\). Moreover, the average distance of a random IoT device from its serving SBS, \(\mathbb{E}[r]\), is derived using the density function (\ref{beta_dist}) as \(\frac{a}{a+b}R_s\). Therefore, the mean-field SINR writes:
\begin{equation}
\gamma(t, p, I_{mf}) =\cfrac{p(t,\textbf{X}_t)(2+\eta^2)L_0(\frac{a}{a+b}R_s)^{-\alpha(\frac{a}{a+b}R_s)}}{\mathrm{N}_0B_w+I_{mf}(t)}.
\end{equation}
Finally, the mean-field utility function is expressed as follows:
\begin{equation}
F(t, p, I_{mf})  = \log_2\big (1 + \gamma(t,p, I_{mf}) \big ).
\end{equation}

The mean-field regime describes the mass behaviors of IoT devices in a massive IoT network, allowing the generic player to make the optimal decision by just responding to the mean-field. By expressing inter-cell interference in terms of an expectation over the mean-field that changes with time according to the FPK equation, we achieve a remarkable degree of economy in the description of population dynamics.
\begin{prop}
The finite mean-field interference is derived for \(t \in [0,T]\) as follows:

\begin{align}
I_{mf}(t) & =  2\sigma_2^2 \int\limits_0^{2\pi}\int\limits_{R_{min}}^{R} \Lambda_0(r,\theta)r^{-\alpha(r)} r dr d\theta \cdot \nonumber\\ &\int_{\Omega}p(t,\textbf{x})m(t,\textbf{x})d\textbf{x},
\end{align}

where \(R_{min}=\max\left(R_{safe},\frac{1}{2\sqrt{\beta_s}}-R_s\right)\) and \(R_{safe}\) represents the minimal distance between a SBS and the nearest interfering IoT device.\\
\end{prop}
\begin{IEEEproof}
Without loss of generality, we derive the finite mean-field interference \(I_{mf}\) for a generic player whose generic SBS is at the origin.\\
The mean-field interference in \(\mathcal{C}_R\) for a given time instant \(t \in [0,T]\) is expressed as:
\begin{equation}
I_{mf}(t) =  \mathbb{E}\left [ \sum_{j=1}^{|\mathcal{N}_{R,0}|}p(t,\textbf{X}_{j,t})H^2_{j}(t)L_{j}(r) \right ].
\end{equation}
Since the transmit powers of the interferes are independent of the point process, we write
\begin{equation}
I_{mf}(t) =  \mathbb{E}[p(t,\textbf{X}_t)] \mathbb{E}[H^2_{outdoor}]\mathbb{E}\left [ \sum_{j=1}^{|\mathcal{N}_{R,0}|} L_{j}(r) \right ],
\end{equation}
where 
\begin{equation}
\mathbb{E}[p(t,\textbf{X}_t)] = \int_{\Omega}p(t,\textbf{x})m(t,\textbf{x})d\textbf{x},
\end{equation}
and
\begin{equation}
\mathbb{E}[H^2_{outdoor}] =  2\sigma_2^2.
\end{equation}
It is worth noting that \(H_{outdoor}\sim Rayleigh(\sigma_2)\), thus, \(H^2_{outdoor}\) follows an exponential distribution with parameter \(1/2\sigma_2^2\). \\
Moreover, since the nearest neighbor distance in a PPP is Rayleigh distributed \cite{t1}. Thus, the average distance between a SBS and a cell-edge interfering IoT device can be written as \(\frac{1}{2\sqrt{\beta_s}} - R_s\) if \(\beta_s \leq \frac{1}{4R_s^2}\). Therefore, the distance between the SBS and the nearest interfering IoT device is computed as \(R_{min}=\max(R_{safe},\frac{1}{2\sqrt{\beta_s}}-R_s)\), where \(R_{safe}\) represents the radius of the SBS's safety zone, i.e., the minimal distance between a SBS and the nearest interfering IoT device.\\
Then, by using Campbell's formula \cite{t1}, we write:
\begin{equation}
\mathbb{E}\left [ \sum_{j=1}^{|\mathcal{N}_{R,0}|} L_{k,j}(r) \right ] = \int\limits_0^{2\pi}\int\limits_{R_{min}}^{R} \Lambda_0(r,\theta)r^{-\alpha(r)} r dr d\theta.
\end{equation}
This concludes the proof.

\end{IEEEproof}
\subsection{Stationary Regime}
In the stationary mean-field regime \cite{s1,s2,c24}, the following additional assumptions are made:\\
- Time homogeneous strategy: All IoT devices follow a time-homogeneous power allocation strategy \(p_t = p : \Omega \to [0,P_{max}]\). \\
- Stationary mean-field: When all the IoT devices follow a time homogeneous power allocation strategy, the mean-field \((m_t)_{t\geq 0}\) converges almost surely to a constant limit \(m\), which we call the stationary mean-field. \\
- Player’s performance: IoT devices evaluate their performance by assuming that the population is infinite and that the corresponding mean-field takes its stationary value at all times. \\
We now turn to the question of the stationary setting. We consider here a stationary FPK equation in which new players randomly arrive in the game according to a Poisson process with intensity \(\frac{\lambda}{c}\) and that the initial state of a new arriving player is randomly drawn from a distribution \(m_s\). The stationary mean-field \(m\) is modeled by a generalized FPK equation expressed as \cite{c22}:
\begin{equation}
-\partial_{e}(pm)(\textbf{x}) - R^{tr}\partial_bm(\textbf{x}) =\frac{\lambda}{c} m_s(\textbf{x}).
\label{FPK_stat}
\end{equation}
Furthermore, with these assumptions, the transmission times (service times) of the IoT devices are independent and identically distributed random variables. Thus, we can model the traffic of IoT devices associated with a SBS at the location \(\textbf{z}\) and using tone \(J\) for transmission by an \(M/G/1\) queue \cite{c21}, where the arrivals are generated by a Poisson process with intensity \(\frac{\lambda_s(\textbf{z})}{c}\), and the average service intensity is given by \(\frac{1}{\mathbb{E}[T^{tr}]}\). Therefore, with this \(M/G/1\) approximation, the probability that a SBS at the location \(\textbf{z}\) has an active IoT device using the tone \(J\) for transmission \(P_{a}(t,\textbf{z})\) when \(t \to \infty\) can be seen as the steady state probability of having at least one customer in the system, expressed as: \\
\begin{equation}
P_a(\textbf{z}) = \frac{\lambda_s(\textbf{z})}{c} \mathbb{E}[T^{tr}].
\end{equation}
Thus, the stationary mean-field interference writes

\begin{align}
I_{mf} & =  2\sigma_2^2 \int\limits_0^{2\pi}\int\limits_{R_{min}}^{R} \Lambda(r,\theta)r^{-\alpha(r)} r dr d\theta \cdot \nonumber\\ &\int_{\Omega}p(\textbf{x})m(\textbf{x})d\textbf{x},
\label{MFI}
\end{align}
where $\Lambda(\textbf{y}) = \beta_s \frac{ \mathbb{E}[T^{tr}]}{c}\int_{ \mathcal{C}_{\textbf{y} } }  \lambda_s(\textbf{z}) f(\textbf{y}|\textbf{z}) d\textbf{z}$, for all $\textbf{y}$ in $\mathbb{R}^2$.\\
As previously stated, the finite population system is approximated by an infinite one due to the following two major characteristics: \\
- The influence of a single player is negligible.\\
- The empirical state distribution almost surely converges to the mean-field, which evolves according to the FPK equation. \\
Thus, two consistency criteria replace the strategic interactions between players in the dynamic differential game: the optimal power allocation strategy is the best reaction to the mean-field, and the mean-field is consistent with the optimal strategy. As a result, the dynamic differential game problem is reduced to a single interaction between the generic player and the mean-field, implying that examining one generic player is sufficient to determine the mean-field equilibrium. This approximation, however, is only significant if the associated approximation error is small. It has been shown that under appropriate conditions, the mean-field regime realizes an \(\epsilon\)-Nash equilibrium for the dynamic differential game, with \(\epsilon\) converging to zero as the number of players goes to infinity \cite{c24}.

\subsection{Mean Field Optimal Control }
In the stationary regime, the utility function of a generic player is given by
\begin{equation}
U(p,I_{mf}) = \theta \int_{\Omega} -F(p(\textbf{x}),I_{mf}) m(\textbf{x})d\textbf{x},
\end{equation}
where $\theta$ is exit time ($\theta \leq T^{tr}$).\\
Thus, the stationary mean-field optimal control problem of a generic player is derived based on (\ref{Opti_p}) and consists in finding \(p^{*}\) and $m^*$ satisfying:  
\begin{equation}
\inf_{p}  \int_{\Omega} -F(p(\textbf{x}),I^*_{mf}) m(\textbf{x}) d\textbf{x},
\label{M_prob}
\end{equation} 
where $m$ is a solution of
\begin{equation}
-\partial_{e}(pm)(\textbf{x}) - R^{tr}\partial_bm(\textbf{x}) =\frac{\lambda}{c} m_s(\textbf{x}), \quad  \textbf{x} \in \Omega,
\label{C1}
\end{equation}
with the following Dirichlet boundary condition
\begin{equation}
m(e,B_{max}) = m(E_{max},b)=0, \quad \forall (e,b) \in \Omega.
\label{C2}
\end{equation}
Note that $I^*_{mf}$ is the mean-field interference at the equilibrium, by assuming that the interfering IoT devices use their optimal power allocation strategy, expressed as:
\begin{align}
I^*_{mf} & =  2\sigma_2^2 \int\limits_0^{2\pi}\int\limits_{R_{min}}^{R} \Lambda(r,\theta)r^{-\alpha(r)} r dr d\theta \cdot \nonumber\\ &\int_{\Omega}p^*(\textbf{x})m^*(\textbf{x})d\textbf{x},
\label{MFI_equi}
\end{align}
Since there is no closed-form solution to the optimization problem ($\ref{M_prob}$) under the partial differential equation constraint ($\ref{C1}$), we will solve it numerically. To accomplish this, the procedures (a) "First discretize then optimize" or (b) "First optimize then discretize" are typically used. In the case of (a), as the name suggests, the optimization problem is discretized first, followed by the development of an iterative approach for the resulting finite dimensional problem. In contrast, the approach (b) necessitates first specifying the optimality conditions at the continuous level, then constructing an iterative approach, and finally discretizing. It is not always the case that one is better than the other. However, a decision must be made to ensure that the discrete system follows the structure of the continuous problem. In this paper, we employed the procedure (b), and since \(F\) is concave in \(p\), the mean-field optimal control is a convex optimization problem. Therefore, the first-order conditions are necessary and sufficient for the optimal solution.

\begin{itemize}
\item \textbf{Optimality Conditions:}
\end{itemize}
The first-order optimality conditions of the mean-field optimal control problem are derived using the adjoint method. Note that even though this approach is used formally in the following, it can be made rigorous. We refer the interested reader to $\cite{c27}$ for a rigorous derivation of these first-order optimality conditions in the time-dependent case. \\ 
Let's start by defining the Lagrangian of the minimization problem (\ref{M_prob}) under the constraint (\ref{C1}) as 
\begin{equation}
\begin{split}
& \mathcal{L}(p,m,\mu) = \int_{\Omega} \bigg ( -F(p(\textbf{x}),I_{mf}^*)m(\textbf{x})+\\
& \mu(\textbf{x}) \big (\partial_{e}(pm)(\textbf{x})  + R^{tr}\partial_bm(\textbf{x})+\frac{\lambda}{c} m_s(\textbf{x}) \big ) \bigg )d\textbf{x},
\end{split}
\label{Lagrange}
\end{equation}
where \(\mu(\textbf{x})\) represent the Lagrange multiplier. \\
By using the integration by parts formula in ($\ref{Lagrange}$), and taking $\mu(e,0) = \mu(0,b)=0$, we get
\begin{equation}
\begin{split}
& \mathcal{L}(p,m,\mu) = -\int_{\Omega} F(p,I_{mf}^*)md\textbf{x} -\int_{\Omega} pm\partial_e\mu d\textbf{x}\\
& -R^{tr}\int_{\Omega}m\partial_b \mu d\textbf{x} + \frac{\lambda}{c} \int_{\Omega} \mu m_sd\textbf{x},
\end{split}
\end{equation}
where we omit the dependency on $\textbf{x}$.\\
Deriving with respect to $p$ and $m$, we obtain the functional derivatives
of the Lagrangian expressed as
\begin{equation}
\partial_p\mathcal{L} (p,m,\mu) = -\partial_pF (p,I_{mf}^*)-\partial_e\mu,
\end{equation}
\begin{equation}
\partial_m\mathcal{L}(p,m,\mu) = -F(p, I^*_{mf})-p\partial_e\mu - R^{tr}\partial_b\mu.
\end{equation}
Let $(p^*,m^*,\mu^*)$ be the optimal solution, then we have 
\begin{equation}
\partial_p\mathcal{L} (p^*,m^*,\mu^*) = \partial_m\mathcal{L} (p^*,m^*,\mu^*) = 0.
\end{equation}
Thus, the optimality conditions are expressed as (\ref{C1}), (\ref{C2}) together with:
\begin{align}
-\partial_pF(p^*,I^*_{mf})-\partial_e\mu^*=0,&
\label{C3}\\
-F(p^*, I^*_{mf})-p^*\partial_e\mu^* - R^{tr}\partial_b\mu^*  = 0,\label{C4}
\end{align}
with the following boundary conditions
\begin{equation}
\mu^*(e,0) = \mu^*(0,b)=0.
\label{C5}
\end{equation}
 Since the mean-field represents a probability density it is natural to supplement the optimality conditions with the further condition
 \begin{equation}
 \int_{\Omega}m^*(\textbf{x})d\textbf{x} = 1.
 \label{C6}
 \end{equation}
Finally, it is worth noting that the equation ($\ref{C4}$) reflects the adjoint equation of ($\ref{C1}$), popularly known as the Hamilton-Jacobi-Bellman equation in mean-field game theory.
\subsection{The Algorithm}
The coupled optimality conditions are solved iteratively until the convergence point is reached to achieve the stationary mean-field equilibrium. We use a successive sweep method consisting of generating a sequence of nominal solutions $p_0, p_1,\dots, p_k, \dots$ that converges to the optimal power allocation strategy $p^*$. This iterative approach, which has proven to be effective in $\cite{c23}$, is summarized in $\textbf{Algorithm 1}$ in our context. Moreover, a gradient-based implementation of this approach can be found in $\cite{c25}$. Finally, we refer the reader to $\cite{c26}$ for a review of several aspects of numerical approaches for mean-field control problems.
\begin{algorithm}
\kwInit{}
Generate random vector \( p_{0}\) \;\vspace{0.2cm}
 \kwLearning{}
Find \(m\) using (\ref{C1}) with boundary conditions (\ref{C2})\;
Normalize \(m\) to verify the conditon (\ref{C6})\;
Estimate mean-field interference $I_{mf}$ using (\ref{MFI_equi})\;
Find \(\mu\) using (\ref{C4}) with boundary conditions (\ref{C5})\;
 Update transmit power \(p\) using (\ref{C3})\;
 Repeat until convergence : go to line 2\;
 \caption{Stationary Mean-Field~Equilibrium}
\end{algorithm}
\section{Numerical Investigations}
In this section, we present numerical results to evaluate the proposed algorithm under spatiotemporal dynamic. We offer some explanations below on how to numerically solve the \textbf{Algorithm 1} using a finite difference method. It is worth noting that the suggested algorithm improves in efficiency as the network spatiotemporal density increases. It means that the computational complexity remains constant regardless of the SBS density or the IoT traffic, and the number of iterations required to reach the optimal power allocation strategy is bounded within tens of iterations, see Figure 1.

\begin{figure}[h]
      \centering
      \includegraphics[width=\linewidth]{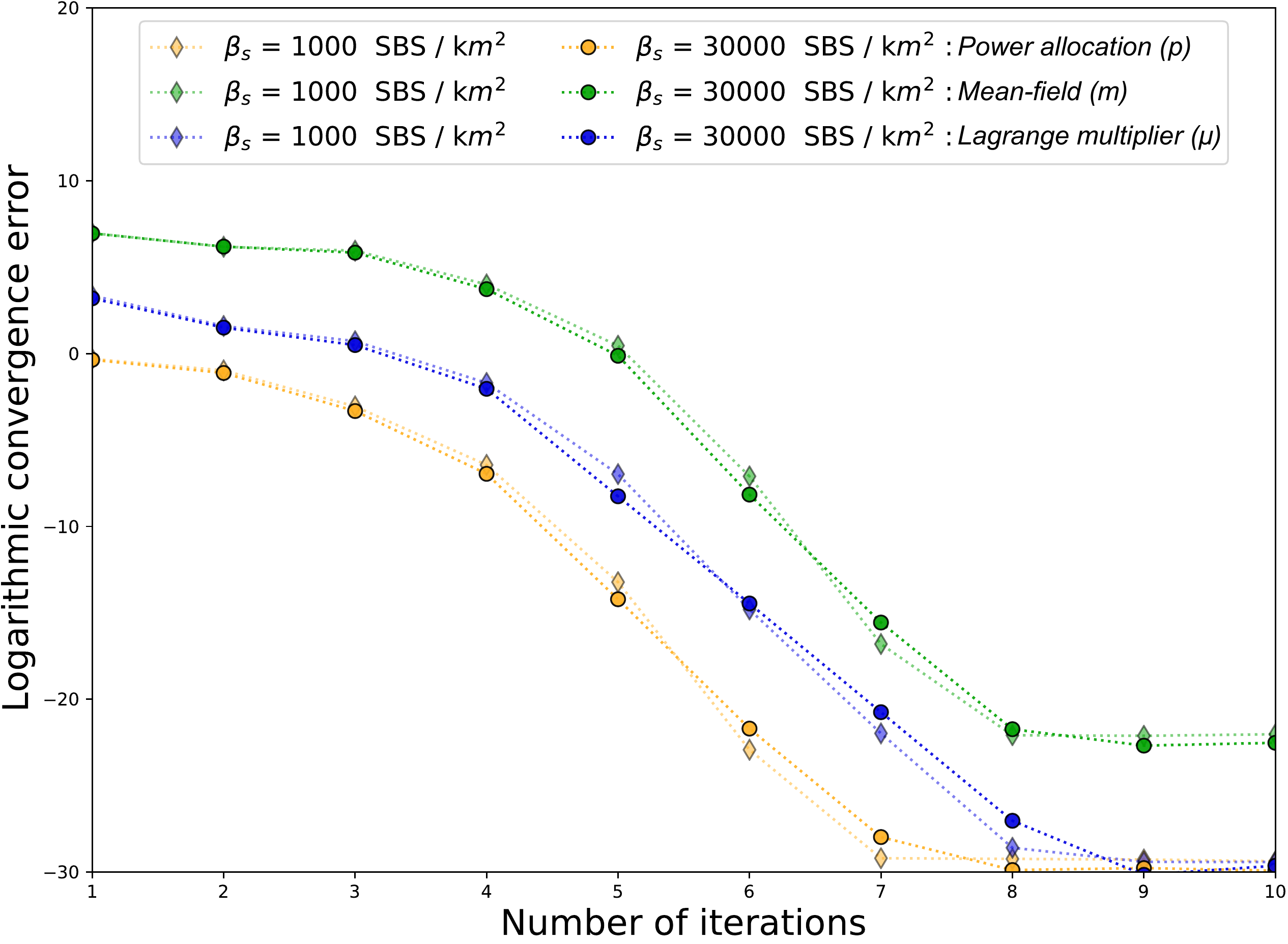}
      \caption{Logarithmic Convergence Error of the iterative $\textbf{Algorithm 1}$ (power allocation, mean-field, Lagrange multiplier) as a function of number of iterations for two different SBS densities and for the same starting point $p_0(e,b) = P_{max}$ for all $(e,b) \in \Omega$}. 
      \label{erreur}
\end{figure}

%****************************************************************************%
\begin{figure*}[!t]
\centering
\begin{subfigure}[t]{0.32\textwidth}
         \centering
         \includegraphics[width=\textwidth]{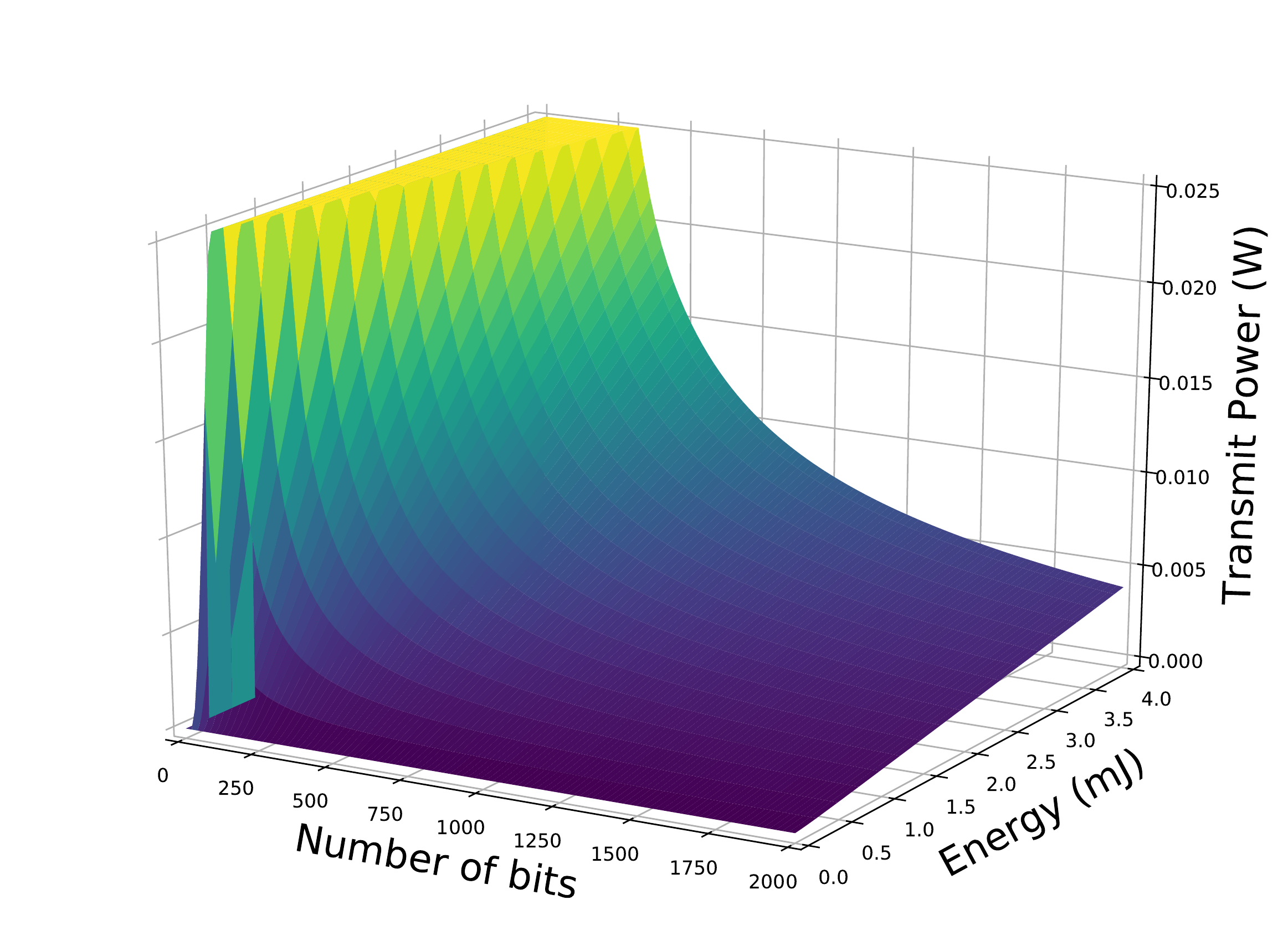}
         \caption{MCS level 0}
\end{subfigure}
\begin{subfigure}[t]{0.32\textwidth}
         \centering
         \includegraphics[width=\textwidth]{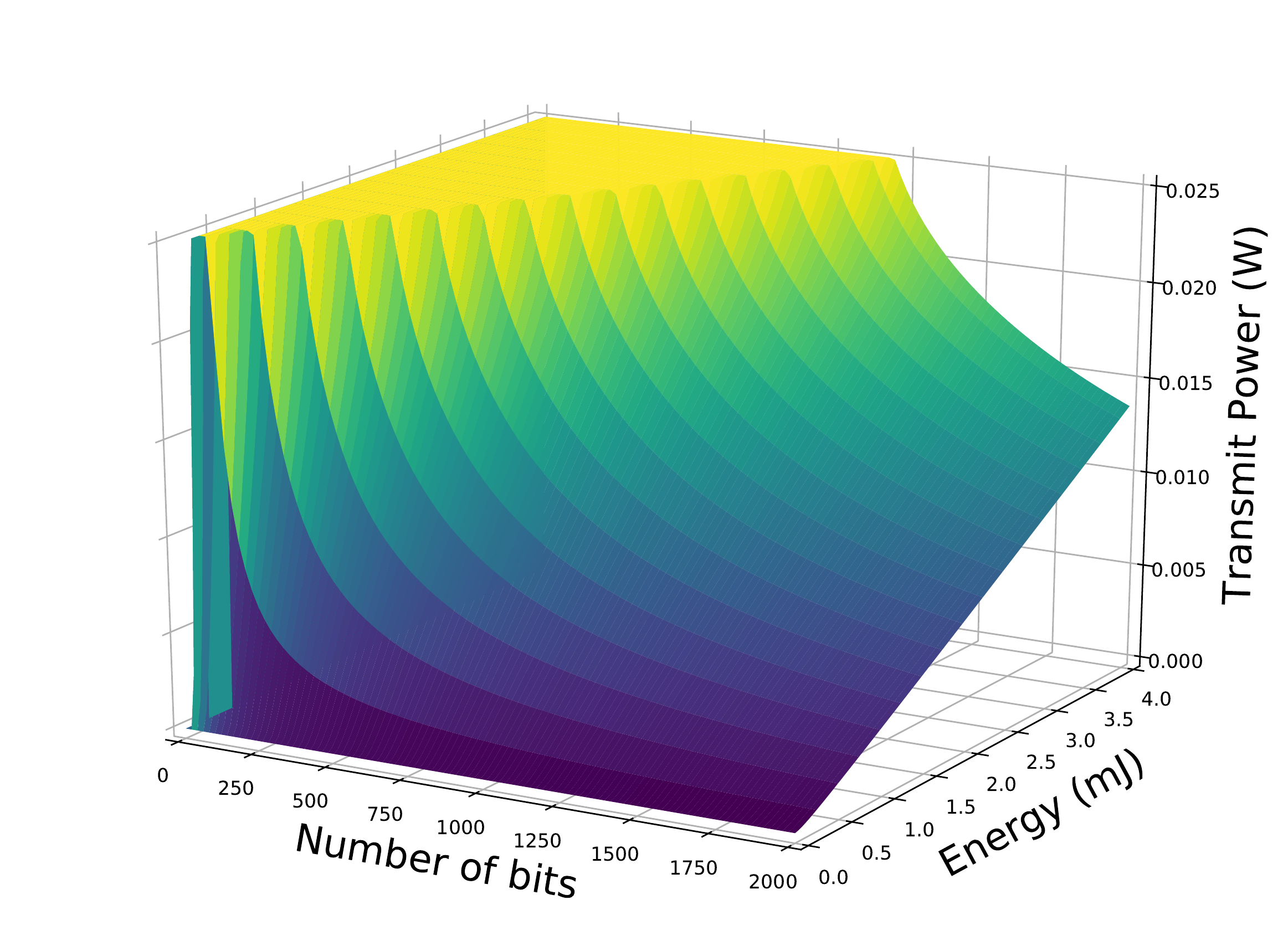}
         \caption{MCS level 4}
\end{subfigure}
\begin{subfigure}[t]{0.32\textwidth}
         \centering
         \includegraphics[width=\textwidth]{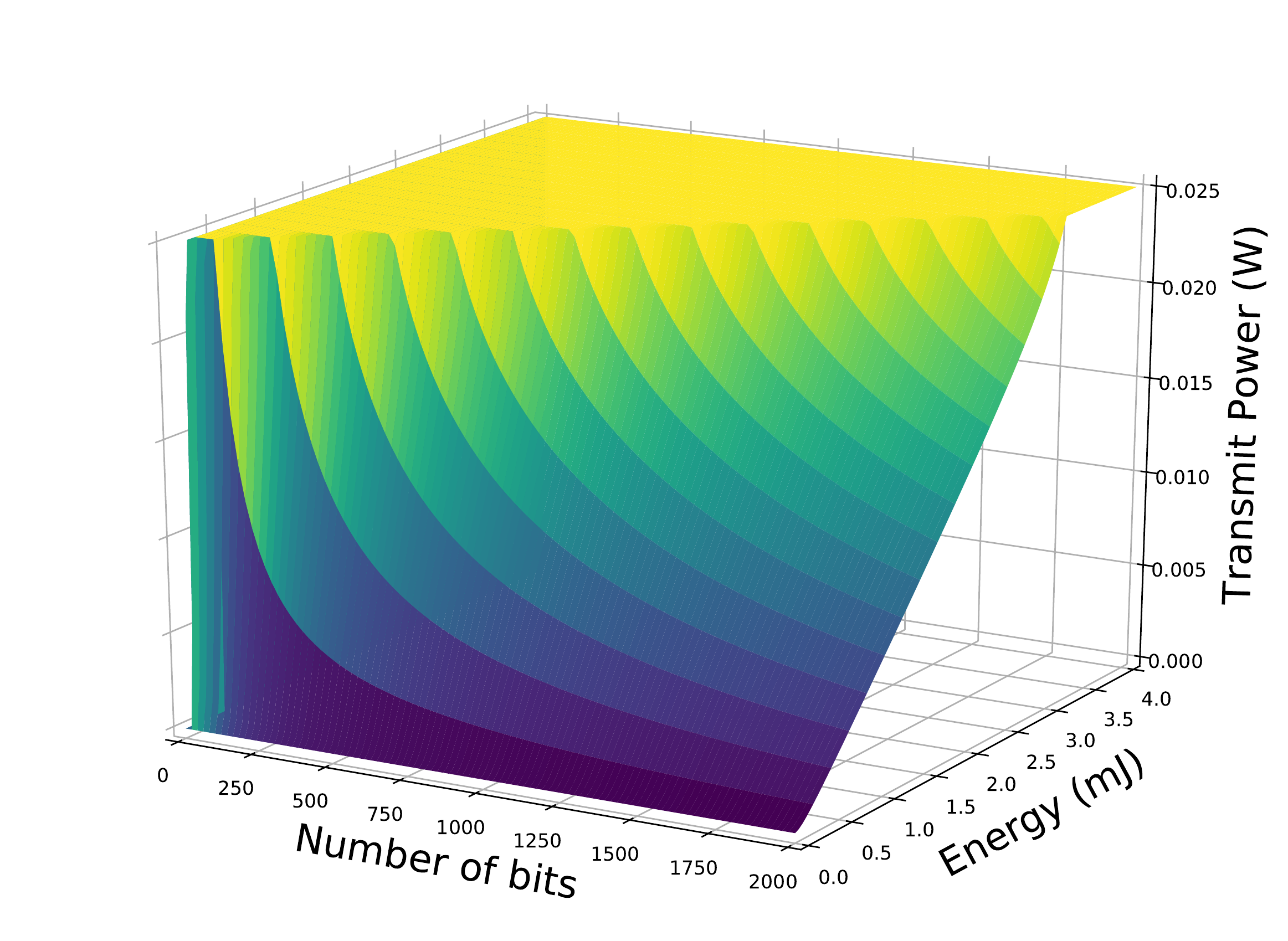}
         \caption{MCS level 8}
\end{subfigure}
\caption{Optimal power allocation strategy as a function of energy and bit numbers for various MCS levels.}
\label{fig:fig}
\end{figure*}
%*************************************************************************%
\subsection{Finite Difference Method}
To numerically solve the algorithm 1, we work on a discretized grid in \(\Omega\). Let us consider two positive integers, \(X\) and \(Y\). We define the space steps by $\delta E=\frac{E_{max}}{X}$, $\delta B = \frac{B_{max}}{Y}$, and for \(i = 0,\dots,X\), \( j= 0,\dots,Y\), we denote \(f_{i,j}\) the numerical approximations of \(f(i\delta E, j \delta B)\) for any function \(f\). The FPK equation (\ref{C1}) is computed iteratively using a upwind-type finite difference scheme by:
\begin{equation}
\begin{split}
& -\frac{p_{i+1,j}m_{i+1,j} - p_{i,j}m_{i,j}}{\delta E} -\\
& R^{tr}\frac{m_{i,j+1}-  m_{i,j}}{\delta B}  = \frac{\lambda}{c} (m_s)_{i,j}.
\end{split}
\end{equation}
Moreover, for an arbitrary point \((i,j)\), the optimality conditions (\ref{C3}), (\ref{C4}) are discretized as follows:
\begin{equation}
 -F_{i,j} - p_{i,j}\frac{\mu_{i,j} - \mu_{i-1,j}}{\delta E} - R^{tr} \frac{\mu_{i,j} - \mu_{i,j-1}}{\delta B}  = 0,
\end{equation}
\begin{equation}
 - \frac{\partial F_{i,j}}{\partial p_{i,j}} - \frac{\mu_{i,j}-\mu_{i-1,j}}{\delta E} = 0,
  \end{equation}
where \(F_{i,j}\) denote the discretized mean-field utility function given as:
\begin{equation}
\begin{split}
F_{i,j} = \log_2\left(1 + \frac{p_{i,j}(2+\eta^2)L_0\left(\frac{a}{a+b}R_s\right)^{-\alpha\left(\frac{a}{a+b}R_s\right)}}{\mathrm{N}_0 B_w+ I_{mf}}\right).
\end{split}
\end{equation}
The fundamental idea underlying the finite difference method is to replace continuous derivatives with difference formulas that only involve discrete values related to grid points. The key parameters of the grid are $\delta E$ and $\delta B$, and when these parameters approach zero, any numerical solution will approach the true solution to the original differential equation. As a result, we chose $X = 50$ and $Y = 200$ in the numerical simulation since $E_{max} \ll B_{max}$.
\subsection{Numerical Analysis}

\begin{table}
\centering 
\caption{ Simulation parameters }
\begin{tabular}{|p{1.7 cm}|p{1.7 cm}|p{3.1 cm}|}
  \hline
 \centering  \textbf{Parameter} &  \centering \textbf{Values} & \centering \textbf{Description} \tabularnewline 
  \hline
 \centering \(a\) & \centering 2& \centering parameters of the beta\tabularnewline 
    \cline{1-2}
  \centering \(b\) &  \centering 4& \centering distribution  \tabularnewline 
 \hline
 \centering \(R_s\) &  \centering 20 $m$ & \centering small cell radius\tabularnewline 
  \hline
  \centering \(R\) &  \centering 10 k$m$ & \centering network radius\tabularnewline 
  \hline
  \centering \(R_{safe}\) &  \centering 3, 4 $m$ & \centering Minimal distance between a SBS and the nearest interfering IoT device\tabularnewline 
  \hline
  \centering \(\beta_s\) &  \centering \(10^3\) \(\to\) \(3 \times 10^5\) SBS / k\(m^2\)& \centering small cell density \tabularnewline 
  \hline
  \centering \(\beta_u\) &  \centering 500 & \centering average number of IoT device per small cell\tabularnewline
  \hline
  \centering \(\lambda_u\) &  \centering\( \frac{1}{60}\), \(\frac{1}{300}\), \(\frac{1}{900}\), \(\frac{1}{3600}\), \(\frac{1}{10800} \) \\ packet / s & \centering average IoT devices arrival rate \tabularnewline 
  \hline
  \centering \(\lambda_s\) &  \centering\(  \beta_u \lambda_u\) & \centering IoT traffic per small cell \tabularnewline 
  \hline
  \centering \(\lambda\) &  \centering\(  \beta_s\pi R^2  \lambda_s \) & \centering IoT traffic in the network\tabularnewline 
   \hline
  \centering \(P_{max}\) &  \centering 0.025 W \\  ($\approx$ 14 dBm) & \centering Maximal transmit power \tabularnewline 
  \hline
   \centering \(B_{max}\) &  \centering 2 kbits \\ (= 2000 bits) & \centering The boundary of the state \\ domain\tabularnewline 
  \cline{1-2}
   \centering \(E_{max}\) &  \centering 0.004 J & \tabularnewline 
  \hline
  \centering \(\mathrm{N}_0\) &  \centering -174 dBm/Hz & \centering Thermal noise density \tabularnewline 
  \hline
   \centering \(N^{ru}\) &  \centering 1 & \centering Number of resource units   \tabularnewline 
  \hline
   \centering \(B_w\) &  \centering 15 kHz & \centering Tone bandwith   \tabularnewline 
  \hline
   \(\alpha_1, \alpha_2, \alpha_3, \alpha_4\) &  \centering 2, 3, 4, 6 & \centering the path-loss exponents  \tabularnewline 
  \hline
\end{tabular}
\end{table}
In this simulation, we consider that the usual information bits generated by IoT devices are around \(600-1200\) bits \cite{R6}, and that each packet contains a header of approximately \(520\) bits. As a result, each packet transmitted by an IoT device is generally \(680-1840\) bits long. Furthermore, we also consider that an IoT device selects an energy budget from \(1\) mJ to \(3\) mJ to transmit those bits. Therefore, we assume that the packet size in kbits (resp. the energy budget in J) of an IoT device is randomly drawn from a truncated normal distribution \(\mathcal{N}( 1,  0.3^2) \) (resp. \(\mathcal{N}(0.002, 0.001^2)\)) on \([0.6, 1.9]\) (resp. \([0.001, 0.003]\)) with a probability density function denoted as \(f_B\) (resp. \(f_E\)). Thus, the state distribution of the arriving player \(m_s\) is given as \(f_B \times f_E\). It's worth noting that for a valid Dirichlet boundary condition (\ref{C2}), the boundary of the state domain \(B_{max}\), and \(E_{max}\) must be strictly greater than the maximum possible packet size, i.e., \(1900\) bits, and the maximum permitted energy budget, i.e., \(3\) mJ, respectively. \\
We also assume a homogeneous IoT traffic at each SBS, i.e., \(\lambda_s(\textbf{z}) = \lambda_s\) for all \(\textbf{z} \in \mathbb{R}^2\). Thus, by using Campbell’s formula to derive the average IoT traffic in the network, and Wald's identity to derive the average IoT traffic per small cell, we write
\begin{equation}
\mathbb{E}[\lambda] = \mathbb{E}[\lambda_s] \beta_s \pi R^2 = \lambda_u \beta_u \beta_s \pi R^2,
\label{AverT}
\end{equation}
where \(\lambda_u = \mathbb{E}[\lambda_k]\) represents the IoT devices' average arrival rate.\\
Note that the values of the traffic's parameters will be selected using the equation (\ref{AverT}). Table IV shows the values of the key simulation parameters.\\

\noindent\textbf{Optimal Transmit Power:} Figure 2 illustrates the optimal power allocation strategy as a function of energy and bit numbers for various MCS levels. It is worth noting that no IoT device has an interest in remaining in the game. Indeed, when an IoT device establishes an energy budget for transmission, it will exhaust it completely to optimize its throughput. Additionally, as shown in Figure 2, each IoT device can adjust its transmit power rest on its available energy and remaining bits. And, as the MCS level grows, the number of points in \(\Omega\) for which \(p^*(e, b) = P_{max}\) grows as well. More precisely, since increasing the MCS level results in a shorter transmission time, IoT devices have the potential to send their packets with higher transmit power. \\

\noindent\textbf{Mean-Field Interference:} Aside from determining the optimal power allocation strategy, the proposed approach also allows us to forecast average network interference while accounting for spatial randomness and IoT traffic. Figure 3 shows the mean-field interference as a function of SBS density for various average arrival rates. As expected, the mean-field interference increases as the IoT traffic grows and has a logarithmic shape when the network becomes ultra-dense. Additionally, Figure 4 illustrates the mean-field interference as a function of MCS level for various average arrival rates. We can observe that even though IoT devices communicate with higher transmit power, the average interference in the system decreases when they use a higher MCS level for transmission. This is due to the fact that transmission time decreases, allowing the IoT devices to spend less time in the network and quit the game faster. \\

\begin{figure}[h]
      \centering
      \includegraphics[width=\linewidth]{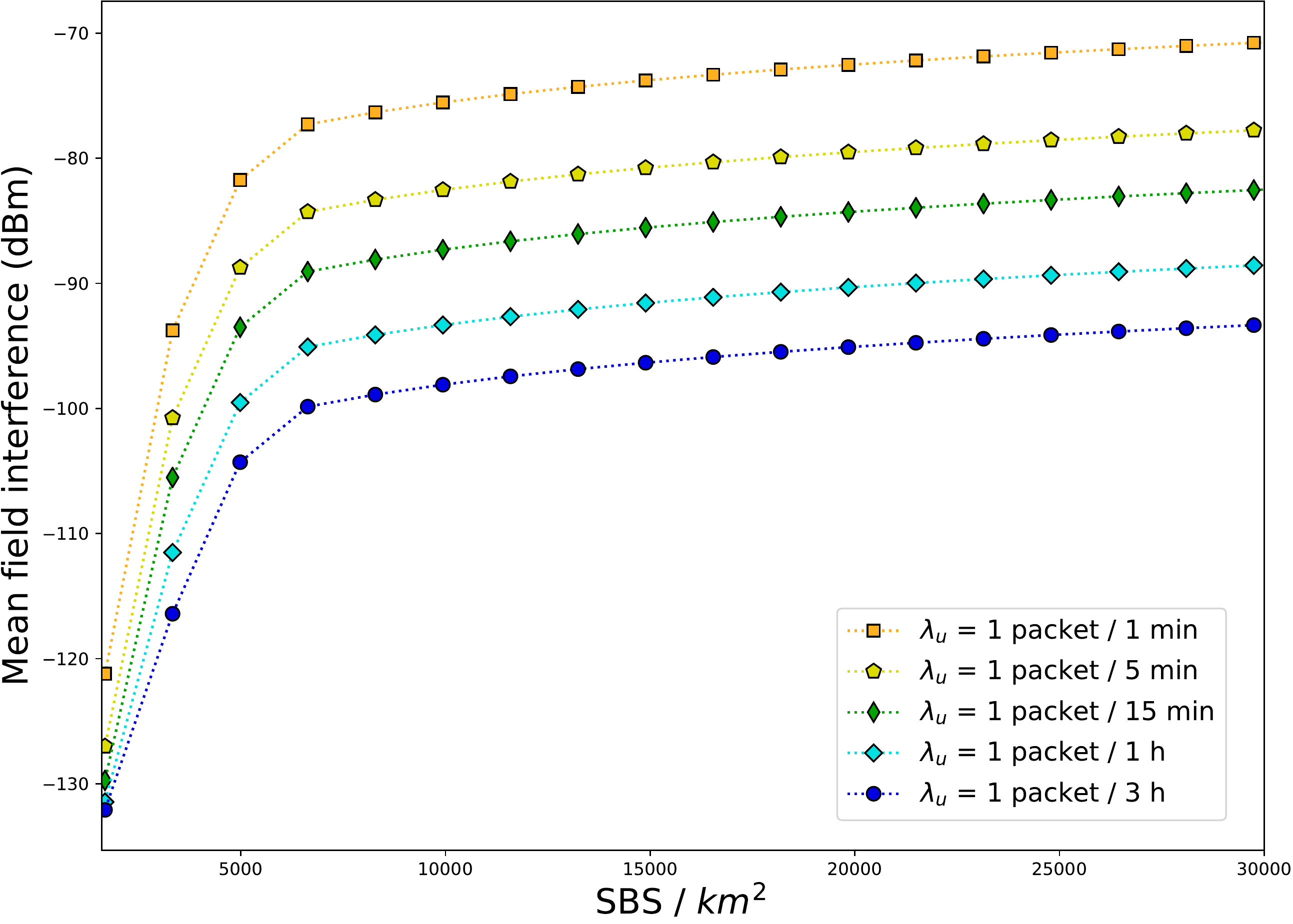}
      \caption{The mean-field interference as a function of SBS density for various arrival rates (MCS level = 8, \(R_{safe} = 4\) m).}
      \label{inter_sbs}
\end{figure}

\begin{figure}[h]
      \centering
      \includegraphics[width=\linewidth]{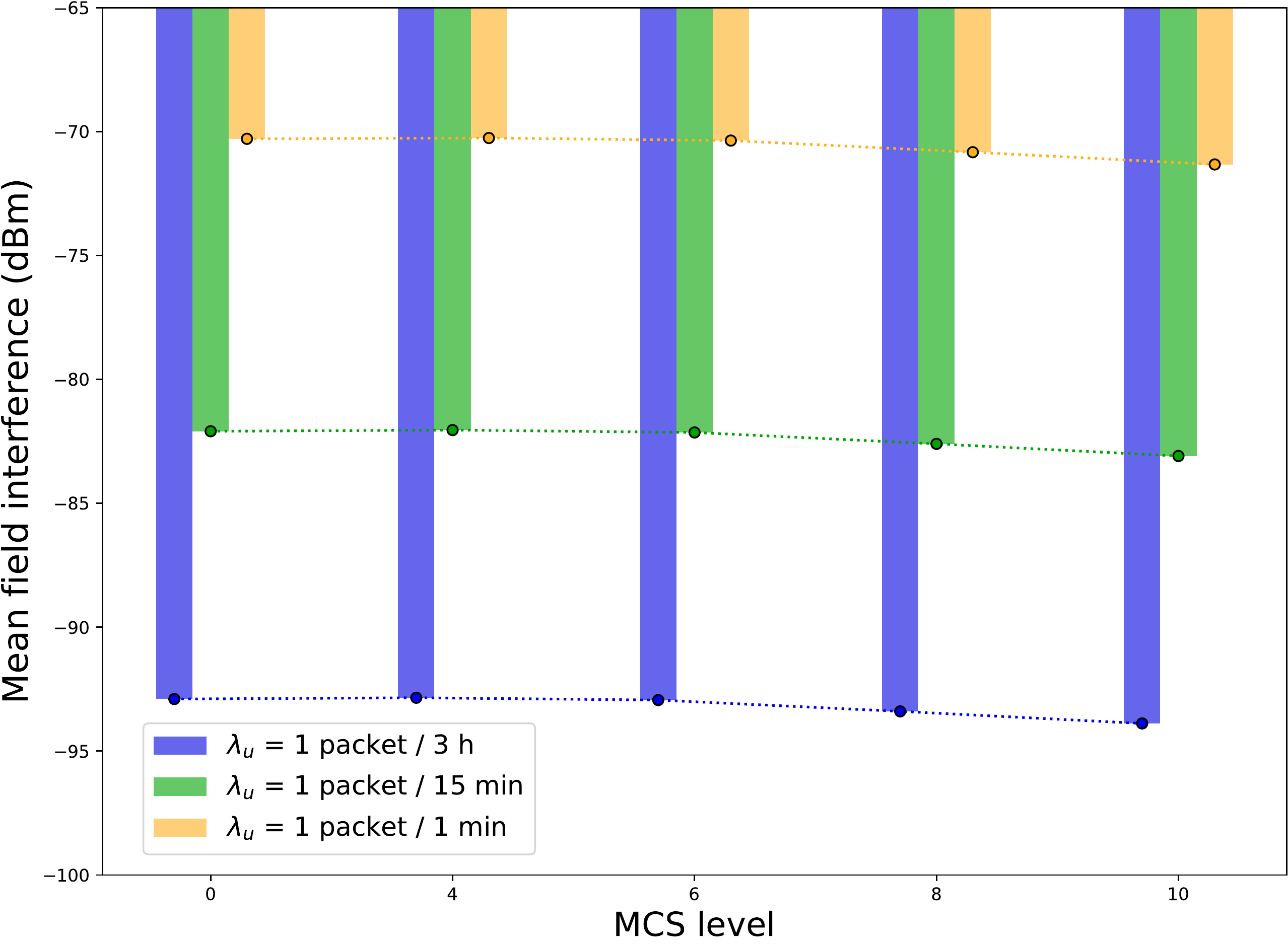}
      \caption{The mean-field interference as a function of MCS level for various arrival rates (\(\beta_s = 30000\) SBS / \(km^2\), \(R_{safe} = 4\) m).}
      \label{inter_tbs}
\end{figure}

\noindent\textbf{Transmission Evaluation and Repetition Protocol:} Figure 4 shows the average SINR as a function of the distance for various energy budgets, packet sizes, and MCS levels. The influence of the MCS level is quite significant, with an approximate \(11\) dB rise in average SINR between the MCS level \(0\) and the MCS level \(8\). Figure 6 also shows the average SINR, but this time, as a function of SBS density for different average arrival rates and safety distances.  It's worth noting that expanding the SBS's safety zone from \(3\) to \(4\) meters increases the average SINR by \(3\) dB. In this simulation, we investigated single-tone transmission without repetition. Since the typical SINR range is between \(10\) and \(70\) dB, as seen in both figures. We may infer that repetitions are unnecessary in an ultra-dense small cell environment, even with heavy traffic, especially if we employ a robust modulation and coding scheme (high MCS level) and enlarge the safety zone. We further validate this observation by plotting the average packet success rate as a function of packet size for various energy budgets and distances under extreme conditions (SBS density of \(300000\) SBS per \(km^2\) and an average arrival rate of 1 packet per min), see figure 7. According to 3GPP, the repetition value is set to ensure high reliability of \(0.99\). We observe that even the tiny proportion of IoT devices at the edge of small cells can assure excellent reliability under these extreme conditions without relying on repetition, simply by enabling them to use a higher energy budget for transmission.\\

\begin{figure}[h]
      \centering
      \includegraphics[width=\linewidth]{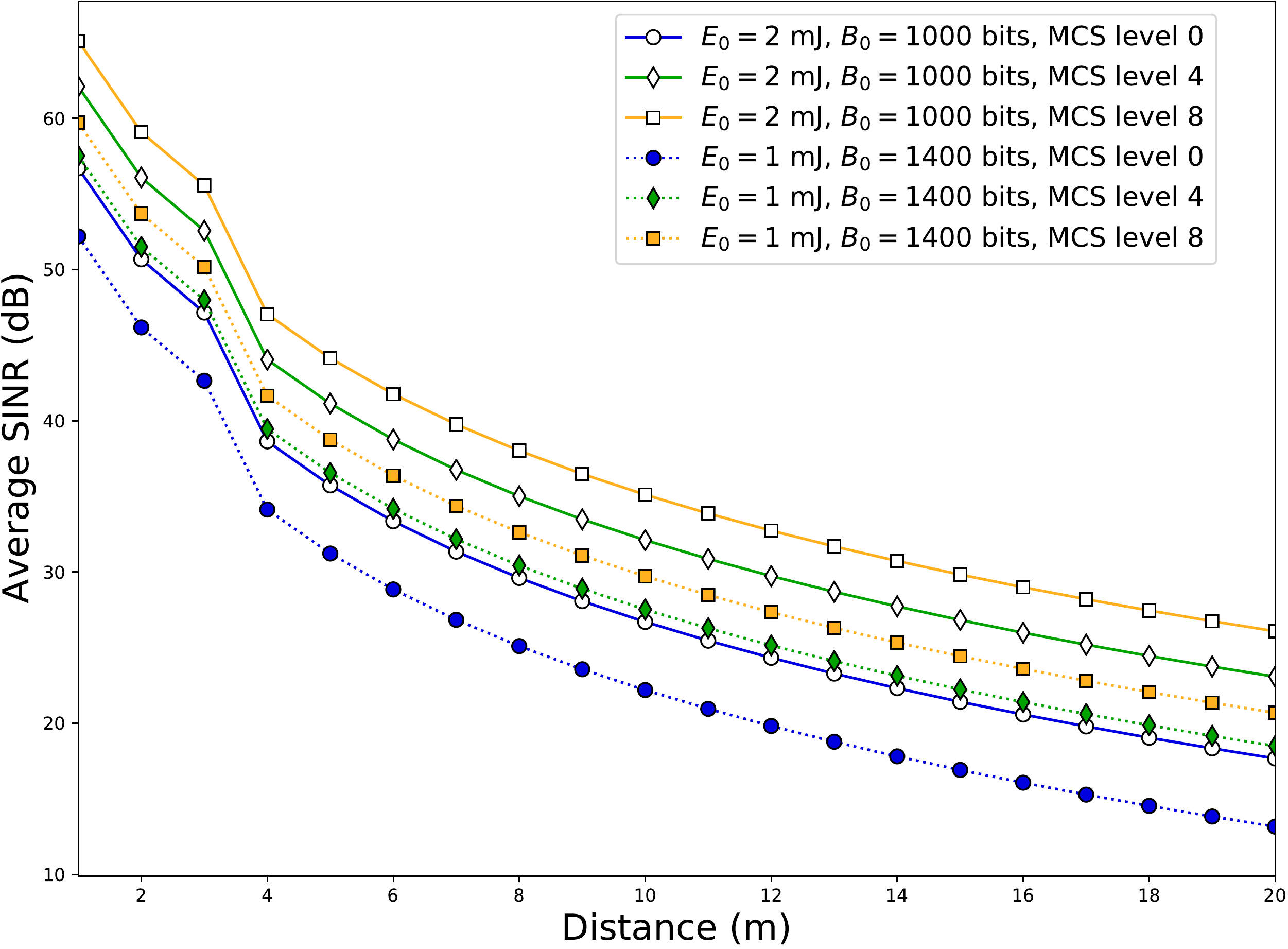}
      \caption{The average SINR as a function of the distance for various energy  budgets, packet sizes, and MCS levels (\(\beta_s = 30000\) SBS / \(km^2\), \(R_{safe} = 4\) m, \(\lambda_u = 1\) packet / \(15\) min).}
      \label{sinr2}
\end{figure}

\begin{figure}[h]
      \centering
      \includegraphics[width=\linewidth]{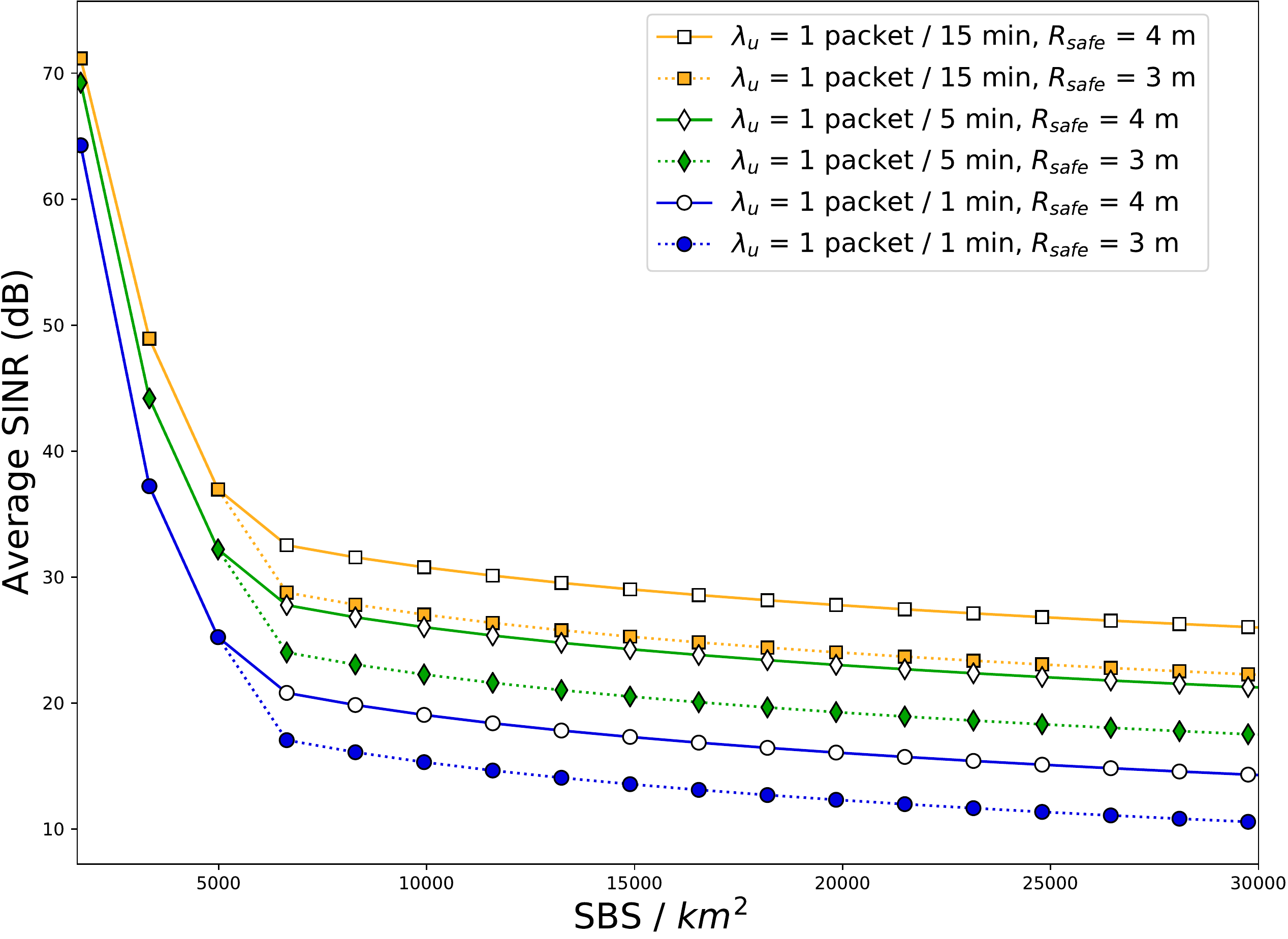}
      \caption{The average SINR as a  function  of  SBS  density  for  different average  arrival  rates  and  safety  distances (MCS level = 8).}
      \label{sinr}
\end{figure}

\begin{figure}[h]
      \centering
      \includegraphics[width=\linewidth]{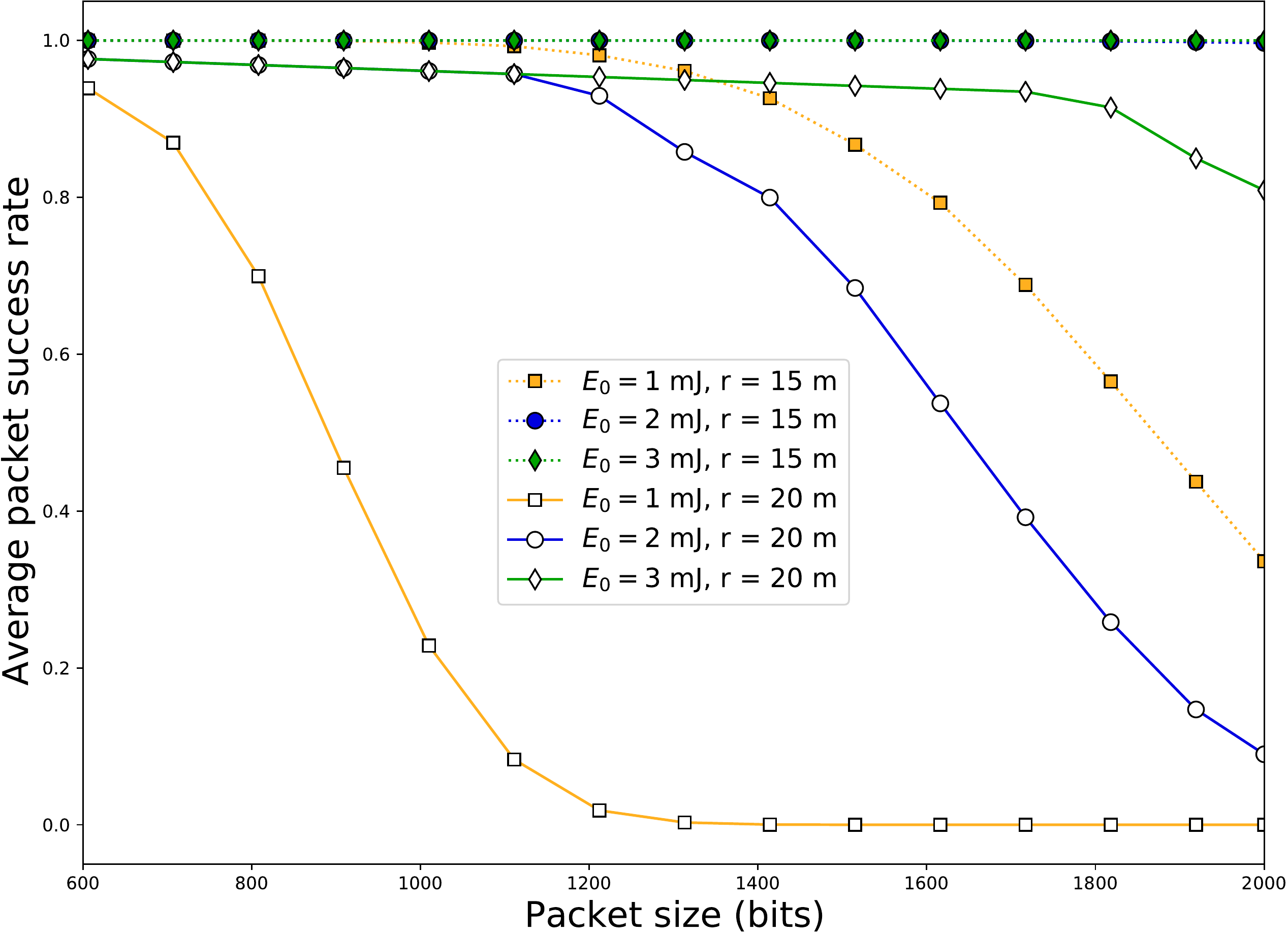}
      \caption{The average  packet  success  rate  as  a  function  of  packet size  for  various  energy  budgets  and  distances (MCS level = 8,  \(R_{safe} = 4\) m, \(\lambda_u\) = 1 packet / 1 min, \(\beta_s\) = \(3 \times 10^5 \) SBS / \(km^2\)).}
      \label{PSR1}
\end{figure}

\section{Conclusion}
We have proposed an traffic-aware distributed power allocation algorithm for ultra-dense small cell NB-IoT networks by leveraging stochastic geometry analysis and mean-field optimal control formulation. The IoT devices in this work are clustered around SBS using a general cluster process, and given their energy budget and packet size, they regulate their transmit power to minimize an average utility function under spatiotemporal fluctuations. Furthermore, the mean-field regime enables IoT devices to distributively compute their power allocation control strategies without having complete awareness of the strategies or states of other IoT devices. The suggested approach also allows us to estimate the average network interference while accounting for spatial randomness and IoT traffic. The simulation results show that the computational complexity of the proposed algorithm remains constant independent of SBS density or IoT traffic, implying that repetitions are unneeded in an ultra-dense small cell environment, even with heavy traffic. Many expansions to the model presented in this paper appear to be feasible and straightforward to compute. We are particularly willing to investigate the uplink performance of the NB-IoT technology in heterogeneous networks where macro users become a dominant source of interference when they share the same PRB as NB-IoT. This would be an interesting topic for future work.

\section*{Acknowledgments}

%%%%%%%%%%%%%%%%%%%%%%%%%%%%%%%%%%%%%%%%%%%%%%%%%%%%%%%%%%%%%%%%%%%%%%%%%%%%%%%%

\addtolength{\textheight}{-1cm} 

%%%%%%%%%%%%%%%%%%%%%%%%%%%%%%%%%%%%%%%%%%%%%%%%%%%%%%%%%%%%%%%%%%%%%%%%%%%%%%%%}

\end{document}